# Kleene Algebra with Domain

JULES DESHARNAIS
Université Laval
and
BERNHARD MÖLLER and GEORG STRUTH
Universität Augsburg

We propose Kleene algebra with domain (KAD), an extension of Kleene algebra with two equational axioms for a domain and a codomain operation, respectively. KAD considerably augments the expressiveness of Kleene algebra, in particular for the specification and analysis of state transition systems. We develop the basic calculus, discuss some related theories and present the most important models of KAD. We demonstrate applicability by two examples: First, an algebraic reconstruction of Noethericity and well-foundedness; second, an algebraic reconstruction of propositional Hoare logic.

Categories and Subject Descriptors: D.2.4 [**Software Engineering**]: Program Verification—*correctness proofs*; F.3.1 [**Logics and Meanings of Programs**]: Specifying and Verifying and Reasoning about Programs—*assertions*; *invariants*; *logics of programs*; *mechanical verification*; *pre- and postconditions*; *specification techniques*; F.3.2 [**Logics and Meanings of Programs**]: Semantics of Programming Languages—*algebraic approaches to semantics*; I.1.3 [**Symbolic and Algebraic Manipulation**]: Languages and Systems—*special-purpose algebraic systems*

## 1. INTRODUCTION

State transition systems are often modelled in a bipartite world in which propositions and actions coexist. Propositions express static properties of states, while actions relate states to model their dynamics. Propositions are usually organized in a Boolean algebra, whereas the sequential, non-deterministic and iterative behaviour of actions is often ruled by a Kleene algebra. Reasoning about state transition systems requires migration between the two parts of the world. This can be modelled by two mappings. One sends actions to propositions in order to express properties of actions. The other sends propositions to actions in order to model propositions as tests, measurements or observations on states, hence as state-preserving actions. This is needed in particular for programming constructs like conditionals or while-loops.

There are two prominent, but complementary realizations of this two-world pic-ture: Propositional dynamic logic (PDL) and its algebraic variants (see, among others, [Harel et al. 2000; Kozen 1979b; Németi 1981; Pratt 1988; 1991; Trnkova and Reiterman 1987]) and Kleene algebra with tests (KAT) [Kozen 1997]. In PDL, only propositions are first-class citizens. This gives the approach a logical flavor. While equivalence of propositions is directly expressible, actions can only be observed indirectly through propositions; the algebra of actions is implicitly defined within that of propositions. However, both mappings are present: modal operators from actions and propositions into propositions and test operators from propositions into actions. This approach is suited for an extensional world, in which actions are completely determined by their input/output behaviour, for instance, when they are modelled as set-theoretic relations. Then, the use of modal operators allows a very versatile and intuitive reasoning. In KAT, only actions are first-class citizens. This gives the approach an algebraic flavour. While equivalence of actions is directly expressible, propositions can only be observed by considering them as actions; the algebra of propositions is embedded as a subalgebra into the algebra of actions, which is a Kleene algebra. Thus only the mapping from propositions to actions is present. The overloading of syntax for propositions and actions leads to particularly economical specifications and proofs. KAT does not make any extensionality assumptions and therefore admits a rich class of models beyond the relational one[1]. Using PDL or KAT, many properties of state transition systems can succinctly be expressed and analyzed. Each approach has its particular advantages and merits. Note, however, that PDL is EXPTIME-complete [Harel et al. 2000], while the equational theory of KAT is PSPACE-complete [Kozen and Smith 1996].

We propose Kleene algebra with domain (KAD) as an extension of KAT and as a reconciliation of KAT and PDL with equal opportunities for propositions and actions. We believe that KAD not only combines the particular advantages of both previous approaches, but also offers additional flexibility and symmetry and yields new structural insights. In particular, beyond this reconciliation, our algebraic abstraction of the domain operation yields a uniform view of hitherto separate approaches to program analysis and development: formalisms based on modal logic such as PDL, formalisms based on algebra such as KAT, set-based formalisms such as B [Abrial 1996] and Z [Spivey 1988], where domain is extensively used, and semantic approaches based on predicate transformers [Dijkstra 1976]. As in KAT, we embed propositions into actions. As in PDL, we also provide a mapping from actions to propositions: the domain operation. Adding domain to KAT is only natural. Relations are the standard model for state transition systems in KAT and PDL. Domain is probably the most natural "modal operator" for relations and KAD supports abstract algebraic reasoning with it. Domain has already been defined algebraically in extensions of Kleene algebra like quantales and relation algebras (cf. [Aarts 1992; Desharnais and Möller 2001; Desharnais et al. 2000; Schmidt and Ströhlein 1993]). But there is no straightforward transfer to KAT. Again, KAD offers several benefits. In opposition to relation algebra, it focuses entirely on the essential operations for state transition systems. Compared with quantales, our approach is entirely first-order and therefore better suited for automated reasoning.

---

[1]Extensionality is studied under the name *separability* in the context of PDL [Kozen 1979a].

Here, our main emphasis is on motivating the definitions, developing the basic calculational aspects and discussing the most interesting models of KAD. We also provide two examples that show its applicability. Many interesting questions, for instance concerning completeness, representability, expressiveness, complexity, the precise relation to modal algebras and a more extensive investigation of applications are postponed to further publications. More precisely, our main results are the following.

—We propose finite equational axiomatizations of a domain and codomain operator for certain idempotent semirings and Kleene algebras.
—We develop a basic domain calculus for KAD. Our axioms capture many natural properties of domain in the relational model and provide new structural insight into the abstract algebraic properties of domain.
—We show that KAD is well-behaved on the standard models of Kleene algebra.
—We define preimage and image operators in KAD. These are very interesting for the specification and analysis of state transition systems and programs.
—We show that Noethericity and well-foundedness are expressible in KAD. We derive properties of these notions.
—We show that KAD subsumes propositional Hoare logic; moreover, we argue that it can serve as the core of abstract axiomatic semantics for imperative programming languages.
—We derive implementation schemata for efficient reachability algorithms for directed graphs within KAD.

Besides these main results there are the following interesting contributions. We show independence of the domain and codomain axioms of KAD. We discuss their compatibility with those for quantales and relation algebra. We provide translations from a class of KAD-expressions to KAT without domain. We introduce two notions of duality that enable a transfer between properties of domain and those of codomain. We show that KAD is not a finitely based variety, whereas all its subalgebras of propositions and all idempotent semirings with domain are.

The remainder of this text is organized as follows. Section 2 introduces idempotent semirings, Kozen's Kleene algebra, some extensions and the standard models for these structures. Section 3 introduces idempotent semirings with tests, KAT and again the standard models. Section 4 presents an equational axiomatization of domain for idempotent semirings. We show independence of the axioms, discuss several extensions and provide some examples for the standard models. Moreover, we outline a basic domain calculus for idempotent semirings. Another important concept, locality of domain and codomain, paves the way to an incorporation of propositional dynamic logic. Section 5 presents two ways of basing an equational axiomatization of codomain for idempotent semirings on that of domain. In Section 6, image and preimage operators are derived from the domain and codomain operators. In Section 7 we derive properties of domain and codomain in KAD. In connection with the Kleene star operator, they allow an abstract treatment of reachability in directed graphs and state transition systems. Section 8 contains some simple metaresults on KAD. Section 9 algebraically reconstructs Noethericity and well-foundedness in KAD. Section 10 shows that propositional Hoare logic is subsumed by KAD. Section 11 draws conclusions and points out further work.

## 2. IDEMPOTENT SEMIRINGS AND KLEENE ALGEBRA

In this section, we introduce idempotent semirings, Kozen's variants of Kleene algebras and certain related structures such as lattice-ordered monoids and quantales. We also present some important models of Kleene algebra, such as the relational model, the language model, the path model, the $(\min, +)$- and $(\max, +)$-models and some of the small finite Kleene algebras of Conway.

Kleene algebras are a class of algebras that axiomatize the regular operations of addition, multiplication and Kleene star as they arise in formal languages and in the analysis of state transition systems and programs. Traditionally, there are two main approaches to Kleene algebra, one based on semirings, the other based on lattices.

### 2.1 Semirings

A *semiring* is a structure $(A, +, \cdot, 0, 1)$ such that $(A, +, 0)$ is a commutative monoid, $(A, \cdot, 1)$ is a monoid, multiplication is left and right distributive with respect to addition and 0 is an annihilator with respect to multiplication ($a \cdot 0 = 0 = 0 \cdot a$).

We call a semiring *trivial* if $0 = 1$, since then for all $a \in A$

$$a = a \cdot 1 = a \cdot 0 = 0,$$

i.e., $A = \{0\}$. Therefore, henceforth we identify semirings with non-trivial semirings.

To abbreviate notation, we write $ab$ instead of $a \cdot b$ and stipulate that multiplication binds stronger than addition.

A semiring is *idempotent* (an *i-semiring*) if addition is idempotent. The class of idempotent semirings is denoted by IS.

The relation $\leq$ defined on an i-semiring $A$ by

$$a \leq b \Leftrightarrow a + b = b \tag{1}$$

for all $a, b \in A$ is a partial ordering, in fact the only partial ordering on $A$ for which $0 \leq a$ for all $a \in A$ such that addition and multiplication are (left and right) monotonic with respect to it. For that reason it is called the *natural ordering* on $A$. By (1), inequalities can be understood as abbreviations for equations. We therefore use the term equation freely for both kinds of expressions.

Obviously, every i-semiring is a semilattice with respect to the natural ordering with least element 0 and addition as join. Thus

$$a \leq c \wedge b \leq c \Leftrightarrow a + b \leq c. \tag{2}$$

In calculations with partial orders, we often appeal to the principles of *indirect inequality* and *indirect equality*. Instead of $a \leq b$ we show $\forall c \,.\, c \leq a \Rightarrow c \leq b$ or $\forall c \,.\, b \leq c \Rightarrow a \leq c$. Likewise, $a = b$ can be proved by showing $\forall c \,.\, c \leq a \Leftrightarrow c \leq b$ or $\forall c \,.\, b \leq c \Leftrightarrow a \leq c$.

## 2.2 Kozen Semirings

A *Kozen semiring* (a K-semiring) [Kozen 1994a] is a structure $(A, +, \cdot, {}^*, 0, 1)$, such that $(A, +, \cdot, 0, 1)$ is an i-semiring, $a^*b$ is the least pre-fixed point of the function $\lambda x.b + ax$ and $ba^*$ is the least pre-fixed point of $\lambda x.b + xa$. Formally, the *Kleene star* $^*$ satisfies the equations

$$1 + aa^* \leq a^*, \qquad (*\text{-}1)$$
$$1 + a^*a \leq a^*, \qquad (*\text{-}2)$$

and the Horn formulas

$$b + ac \leq c \Rightarrow a^*b \leq c, \qquad (*\text{-}3)$$
$$b + ca \leq c \Rightarrow ba^* \leq c, \qquad (*\text{-}4)$$

for all $a, b, c \in A$. The class of K-semirings is denoted by KA.

The expressions $a^*b$ and $ba^*$ are uniquely defined by $(*\text{-}1)$ and $(*\text{-}3)$, and $(*\text{-}2)$ and $(*\text{-}4)$, respectively. We now recall some further standard properties of K-semirings (cf. [Kozen 1994a]). Most of them are also familiar from formal language theory [Eilenberg 1974].

LEMMA 2.1. *Let $A \in$ KA. For all $a, b \in A$,*

$$1 \leq a^*, \qquad (3)$$
$$a^*a^* = a^*, \qquad (4)$$
$$\forall i \in \mathbb{N} . a^i \leq a^*, \qquad (5)$$
$$a^{**} = a^*, \qquad (6)$$
$$(ab)^*a = a(ba)^*, \qquad (7)$$
$$(a+b)^* = a^*(ba^*)^*, \qquad (8)$$
$$a^*b = b + a^*ab = b + aa^*b. \qquad (9)$$

*For all $a, b, c \in A$,*

$$a \leq 1 \Rightarrow a^* = 1, \qquad (10)$$
$$a \leq b \Rightarrow a^* \leq b^*, \qquad (11)$$
$$ac \leq cb \Rightarrow a^*c \leq cb^*, \qquad (12)$$
$$ca \leq bc \Rightarrow ca^* \leq b^*c. \qquad (13)$$

## 2.3 Lattice-Ordered Monoids and Quantales

A *lattice-ordered monoid* (an *l-monoid*) is a structure $(A, +, \sqcap, \cdot, 1)$, such that $(A, +, \sqcap)$ is a lattice, $(A, \cdot, 1)$ is a monoid and left and right multiplication are additive. l-monoids are extensively studied in [Birkhoff 1984]. An l-monoid is *bounded* if it has a least element $0$ and a greatest element $\top$. It is *complete* if the underlying lattice is. A *quantale* [Mulvey 1986] or *standard Kleene algebra* [Conway 1971] is a complete l-monoid in which left and right multiplication is universally additive. Quantales have been investigated in contexts like the logic of quantum mechanics [Mulvey 1986] and algebraic models of certain linear logics [Yetter 1990]. *D-monoids* or *b-monoids* are l-monoids whose lattice reducts are distributive or Boolean, respectively. A *d-quantale* and *b-quantale*, respectively, is a quantale whose lattice reduct is distributive, and Boolean, respectively. Remembering that a Boolean lattice is a complemented distributive lattice, we use the term *Boolean algebra* as a synonym. b-monoids and b-quantales have been studied, for instance, in [Desharnais and Möller 2001; Desharnais et al. 2000]. Also the sequential algebras of [Hoare and von Karger 1995] are particular b-quantales. In quantales, the Knaster-Tarski theorem guarantees that the Kleene star, which is again defined as the least pre-fixed point of a monotonic function, always exists. This is in contrast to KA, where completenes of the underlying semilattice is not assumed.

The main results of this paper are entirely based on i-semirings and not on quantales.

A first reason for this decision is that i-semirings are more general than quantales. Every quantale is an l-monoid; every K-semiring and every l-monoid is an i-semiring. K-semirings are first-order structures whereas, due to completeness, quantales are essentially higher-order. A certain price we have to pay is that without the assumption of completeness we cannot freely use Galois connections as a very elegant means of defining certain functions, as would be the case in quantales.

A second reason is that in b-modules and b-quantales there is a notion of complementation. When reasoning about programs, elements of a b-monoid represent programs as input/output relations. Hence the complement of such an element relates all states that are not in the input/output relation. While this is alright for sequential programs, this concept meets severe difficulties when it comes to parallel programs and IS has the advantage of avoiding this concept.

### 2.4 Example Structures

The classes IS and KA are quite rich. We now present some standard examples. We will later show that the domain and codomain operations are well-behaved on all these structures. In the first examples, we present some of the finite K-semirings with at most 4 elements from Conway's book (cf. [Conway 1971], p.101). We will later use them, in particular, as counterexamples.

EXAMPLE 2.2.

$$\begin{array}{c} 1 \\ | \\ 0 \end{array}$$

*Consider the structure* $A_2 = (\{0,1\}, +, \cdot, 0, 1)$ *with addition and multiplication defined by the tables*

| + | 0 | 1 |   | · | 0 | 1 |
|---|---|---|---|---|---|---|
| 0 | 0 | 1 |   | 0 | 0 | 0 |
| 1 | 1 | 1 |   | 1 | 0 | 1 |

*Then* $A_2$ *is an i-semiring, called the* Boolean semiring, *since* $+$ *and* $\cdot$ *play the roles of disjunction and conjunction.* $A_2$ *can uniquely be extended to a K-semiring by setting* $0^* = 1^* = 1$. □

EXAMPLE 2.3.

$$a - 1 - 0$$

Consider the i-semiring $A_3^1 = (\{a, 0, 1\}, +, \cdot, 0, 1)$ with addition and multiplication defined by the tables

| + | 0 | a | 1 |
|---|---|---|---|
| 0 | 0 | a | 1 |
| a | a | a | a |
| 1 | 1 | a | 1 |

| · | 0 | a | 1 |
|---|---|---|---|
| 0 | 0 | 0 | 0 |
| a | 0 | a | a |
| 1 | 0 | a | 1 |

It can uniquely be extended to a K-semiring by setting $0^* = 1^* = 1$ and $a^* = a$.  □

EXAMPLE 2.4.

$$1 - a - 0$$

Consider the i-semiring $A_3^2 = (\{a, 0, 1\}, +, \cdot, 0, 1)$ with addition and multiplication defined by the tables

| + | 0 | a | 1 |
|---|---|---|---|
| 0 | 0 | a | 1 |
| a | a | a | 1 |
| 1 | 1 | 1 | 1 |

| · | 0 | a | 1 |
|---|---|---|---|
| 0 | 0 | 0 | 0 |
| a | 0 | 0 | a |
| 1 | 0 | a | 1 |

It can uniquely be extended to a K-semiring by setting $a^* = 0^* = 1^* = 1$.  □

EXAMPLE 2.5. Consider the i-semiring $A_3^3 = (\{a, 0, 1\}, +, \cdot, 0, 1)$ which is like $A_3^2$ except for the value of $a \cdot a$:

| + | 0 | a | 1 |
|---|---|---|---|
| 0 | 0 | a | 1 |
| a | a | a | 1 |
| 1 | 1 | 1 | 1 |

| · | 0 | a | 1 |
|---|---|---|---|
| 0 | 0 | 0 | 0 |
| a | 0 | a | a |
| 1 | 0 | a | 1 |

It can uniquely be extended to a K-semiring by setting $a^* = 0^* = 1^* = 1$.  □

EXAMPLE 2.6.

$$b - 1 - a - 0$$

Consider the i-semiring $A_4^1 = (\{a, b, 0, 1\}, +, \cdot, 0, 1)$ with addition and multiplication defined by the tables

| + | 0 | a | 1 | b |   | · | 0 | a | 1 | b |
|---|---|---|---|---|---|---|---|---|---|---|
| 0 | 0 | a | 1 | b |   | 0 | 0 | 0 | 0 | 0 |
| a | a | a | 1 | b |   | a | 0 | 0 | a | a |
| 1 | 1 | 1 | 1 | b |   | 1 | 0 | a | 1 | b |
| b | b | b | b | b |   | b | 0 | a | b | b |

It can be extended to a K-semiring by setting $0^* = a^* = 1^* = 1$ and $b^* = b$. □

There are 18 four-element K-semirings, up to isomorphism.

EXAMPLE 2.7. *Consider a set $A$ and the structure* $\mathsf{REL}(A) = (2^{A \times A}, \cup, \circ, \emptyset, \Delta)$, *where $2^{A \times A}$ denotes the set of binary relations over $A$, $\cup$ denotes set union, $\circ$ denotes relational product, $\emptyset$ denotes the empty relation and $\Delta$ denotes the identity relation $\{(a, a) \mid a \in A\}$.*

*Then $\mathsf{REL}(A)$ is an i-semiring with set inclusion as the natural ordering. It can be extended to a K-semiring by defining $R^*$ as the reflexive transitive closure of $R$ for all $R \in \mathsf{REL}(R)$, that is, $R^* = \bigcup_{i \geq 0} R^i$, where $R^0 = \Delta$ and $R^{i+1} = R \circ R^i$.*

*We call $\mathsf{REL}(A)$ the* relational i-semiring *or* K-semiring *over $A$.* □

EXAMPLE 2.8. *Let $(A, +, \cdot, 0, 1)$ be a semiring and $Q$ be a finite set. Then the set $A^{Q \times Q}$ can be viewed as the set of $|Q| \times |Q|$-matrices with indices in $Q$ and elements in $A$. Now consider the structure $\mathsf{MAT}(Q, A) = (A^{Q \times Q}, +, \cdot, \mathbf{0}, \mathbf{1})$ where $+$ and $\cdot$ are the usual operations of matrix addition and multiplication, and $\mathbf{0}$ and $\mathbf{1}$ are the zero and unit matrices. Then $\mathsf{MAT}(Q, A)$ again forms a semiring, the* matrix semiring *over $Q$ and $A$. $\mathsf{MAT}(Q, A)$ is idempotent if $A$ is. In this case, the natural order is the componentwise order. If the underlying semiring $A$ admits infinite sums, also $Q$ may be infinite.*

*Taking $A$ as the Boolean semiring yields another representation of $\mathsf{REL}(A)$ as $\mathsf{MAT}(Q, A)$ in terms of adjacency matrices.*

*If $A$ is a K-semiring and $Q$ is finite, then $\mathsf{MAT}(Q, A)$ can be extended to a K-semiring (see [Conway 1971]) by partitioning a non-singleton matrix into submatrices $a, b, c, d$, of which $a$ and $d$ are square, and setting*

$$\begin{pmatrix} a & b \\ c & d \end{pmatrix}^* = \begin{pmatrix} f^* & f^*bd^* \\ d^*cf^* & d^* + d^*cf^*bd^* \end{pmatrix},$$

*where $f = a + bd^*c$.* □

EXAMPLE 2.9. *Let $\Sigma^*$ be the set of finite words over some finite alphabet $\Sigma$ and consider the structure $\mathsf{LAN}(\Sigma) = (2^{\Sigma^*}, \cup, ., \emptyset, \{\varepsilon\})$, where $2^{\Sigma^*}$ denotes the set of languages over $\Sigma$, and $\cup$ denotes set union, $L_1 L_2 = \{vw \mid v \in L_1, w \in L_2\}$, where $vw$ denotes concatenation of $v$ and $w$, $\emptyset$ denotes the empty language and $\varepsilon$ denotes the empty word.*

*Then $\mathsf{LAN}(\Sigma)$ is an i-semiring with natural ordering defined by language inclusion. It can be extended to a K-semiring by defining $L^* = \{w_1 w_2 \ldots w_n \mid n \geq 0, w_i \in L\}$.*

*We call $\mathsf{LAN}(\Sigma)$ the* language i-semiring *or* K-semiring *over $\Sigma$. Remember that $\cup$, . and $^*$ are often called* regular operations *and the sets that can be obtained* *from finite subsets of $\Sigma^*$ by a finite number of regular operations are called* regular subsets *or* regular events *of $\Sigma^*$. The equational theory of the rational subsets is also called the* algebra of regular events.

There is a natural homomorphism $L$ from the term algebra over the signature of K-semirings generated by a set $\Sigma$ onto the algebra $\mathsf{REG}(\Sigma)$ of regular subsets of $\Sigma^*$, given by $L(a) = \{a\}$ for each $a \in \Sigma$, $L(a+b) = L(a) \cup L(b)$ and $L(a \cdot b) = L(a).L(b)$. Kozen [Kozen 1994a] has shown that $\mathsf{REG}(\Sigma)$ is the free K-semiring on the generators $\Sigma$. □

EXAMPLE 2.10. *Consider a set $\Sigma$ of vertices (or states). Then subsets of $\Sigma^*$ can be viewed as sets of possible graph paths (or state sequences in a transition system). The partial operation of* join *or* fusion product *of elements of $\Sigma^*$ is defined as*

$$\varepsilon \bowtie \varepsilon = \varepsilon \tag{14}$$

$$\varepsilon \bowtie (y.t) \text{ is undefined}, \tag{15}$$

$$(s.x) \bowtie \varepsilon \text{ is undefined}, \tag{16}$$

$$(s.x) \bowtie (y.t) = \begin{cases} s.x.t & \text{when } x = y, \\ \text{undefined} & \text{otherwise} \end{cases} \tag{17}$$

*for all $s, t \in \Sigma^*$ and $x, y \in \Sigma$. It describes the gluing of paths at a common end point. This operation is extended to subsets of $\Sigma^*$ by*

$$S \bowtie T = \{s \bowtie t \mid s \in S \land t \in T \land s \bowtie t \text{ defined}\}.$$

*Then $\mathsf{PAT}(\Sigma) = (2^{\Sigma^*}, \cup, \bowtie, \emptyset, \Sigma \cup \{\varepsilon\})$ is an i-semiring that we call the* path i-semiring *over $\Sigma$.* □

EXAMPLE 2.11. *Using matrices over the language algebra we can also model labelled transition systems. Assume a set $Q$ of states and a set $\Sigma$ of labels. The matrices in $\mathsf{MAT}(Q, \mathsf{LAN}(\Sigma))$ can be considered as recording possible sequences of labels (traces) that connect two states; if there is no possible transition between two states, the corresponding matrix element is the empty language.* □

EXAMPLE 2.12. *The language example can easily be generalized to an arbitrary monoid $(A, \cdot, 1)$. Then $(2^A, \cup, \cdot, \emptyset, \{1\})$ is the free i-semiring over $(A, \cdot, 1)$. In particular, multiplication (and star) are defined as in Example 2.9.* □

EXAMPLE 2.13. *Set $\mathbb{N}_\infty = \mathbb{N} \cup \{\infty\}$ and define the operations* min *and* + *in the obvious way. Then the structure $(\min, +) = (\mathbb{N}_\infty, \min, +, \infty, 0)$ is an i-semiring, called the* tropical semiring *[Kuich 1997]. Its natural ordering is the converse of the standard ordering on $\mathbb{N}_\infty$. Hence $0$ — the semiring multiplicative unit — is the largest element, so that by (10) $(\min, +)$ can uniquely be extended to a K-semiring by setting $n^* = 0$ for all $n \in \mathbb{N}_\infty$.* □

EXAMPLE 2.14. *Set $\mathbb{N}_{-\infty} = \mathbb{N} \cup \{-\infty\}$ and consider the structure $(\max, +) = (\mathbb{N}_{-\infty}, \max, +, -\infty, 0)$ with operations defined in the obvious way. Then $(\max, +)$ is an i-semiring, called the* max-plus semiring *[Gaubert and Plus 1997]. Its natural ordering coincides with the standard ordering on $\mathbb{N}_{-\infty}$. Unlike the tropical semiring, the max-plus semiring cannot be extended to a K-semiring. The reason is that for $a > 0$ the set $\{a^n \mid n \in \mathbb{N}\} = \{na \mid n \in \mathbb{N}\}$ is unbounded, whereas, according to (5), it should have $a^*$ as an upper bound.* □

## 3. SUBIDENTITIES AND KLEENE ALGEBRA WITH TESTS

We now take the first step towards the definition of domain and codomain operations on IS. We discuss the subidentities of IS, KA and related structures. These are the elements that lie below the multiplicative unit. We also introduce idempotent semirings with tests and Kleene algebras with tests. Finally, we discuss a few important models of these structures.

As a motivation, consider the relational i-semiring from Example 2.7. Here, the domain of a relation is a set. Abstracting to arbitrary i-semirings, the domain operation should be a mapping from the i-semiring to some appropriate Boolean algebra. In the matrix representation for finite relations based on the Boolean semiring, obviously, a *characteristic matrix* can be associated with each set $A$. Setting $n = |A|$, the empty set is characterized by the $n \times n$ zero matrix, the set $A$ by the $n \times n$ unit matrix and all other sets by matrices smaller than the unit matrix. Obviously, there are $2^n$ such matrices, which is also the number of subsets of $A$. Using this abstraction, we model domain and codomain in an i-semiring as an i-semiring endomorphism into the set of elements that are smaller than 1. We now take a closer look at the set of these elements.

### 3.1 Subidentities

An element $a$ of an i-semiring $A$ is a *subidentity* if $a \leq 1$. We denote the set of subidentities of $A$ by $\mathsf{sid}(A)$.

LEMMA 3.1. *The set of subidentities of an i-semiring forms an i-semiring.*

However, this subsemiring is usually too large for our purposes. In the relational i-semiring or in b-monoids, multiplication of subidentities is a meet operation and the set of subidentities is a Boolean sublattice (cf. Section 3.3). In i-semirings, this need not be the case.

LEMMA 3.2. *Multiplication of subidentities in IS is a lower bound operation.*

PROOF. Let $A \in \mathsf{IS}$ and $p, q \in \mathsf{sid}(A)$. Then $p = p1 \geq pq \leq 1q = q$. Thus $pq$ is a lower bound of $p$ and $q$. □ □

LEMMA 3.3. *Multiplication of subidentities in IS (in d-monoids, d-quantales) is not always idempotent.*

PROOF. Consider the i-semiring $A_3^2$ from Example 2.4. Obviously, $a$ is a subidentity that is not multiplicatively idempotent. Since $A_3^2$ is a chain, it is automatically a distributive lattice, hence a d-monoid. Since it is finite, it is automatically complete, hence a d-quantale. The counterexample is minimal for all these structures. □ □

Consequently, multiplication of subidentities in IS is not in general a greatest lower bound operation. Additional properties are required to model sets, propositions or tests in IS.

LEMMA 3.4. *The set of multiplicatively idempotent subidentities of an i-semiring forms a bounded distributive lattice.*

PROOF. Let $\mathsf{isid}(A)$ be the set of multiplicatively idempotent subidentities of $A \in \mathsf{IS}$. We first show that multiplication restricted to $\mathsf{isid}(A)$ coincides with the greatest lower bound operation. By Lemma 3.2 it is a lower bound operation. Let $p, q, r \in \mathsf{isid}(A)$ with $r \leq p$ and $r \leq q$. Then $r = rr \leq pq$, whence $pq$ is the greatest lower bound of $p$ and $q$. Consequently, also

$$pq = qp \qquad (18)$$

for $p, q \in \mathsf{isid}(A)$.

We now show the closure properties of the subalgebra. $0, 1 \in \mathsf{isid}(A)$ is obvious. Let $p, q \in \mathsf{isid}(A)$. Then

$$\begin{aligned}(p + q)(p + q) &= pp + pq + qp + qq \\ &= p + (p \sqcap q) + (q \sqcap p) + q \\ &= p + q\end{aligned}$$

and, using (18),

$$(pq)(pq) = ppqq = pq.$$

We now show that the sublattice is distributive. The first distributivity law

$$p(q + r) = pq + pr$$

holds by the semiring laws. The second distributivity law

$$p + (qr) = (p + q)(p + r)$$

then follows from the first one by lattice algebra.

Finally, the lattice is bounded, since $0, 1 \in \mathsf{isid}(A)$. □ □

Instead of using the set of *all* subidentities, we choose another way (see also the discussion on page 18) that is conceptually much simpler and is introduced in the following subsection.

### 3.2 Test-Semirings and Kleene Algebra with Tests

Following Kozen's approach to Kleene algebra with tests, we say that a *test semiring* (a *t-semiring*) is an i-semiring $A$ with a distinguished Boolean subalgebra $\mathsf{test}(A)$ of $\mathsf{sid}(A)$ with greatest element 1 and least element 0. We call $\mathsf{test}(A)$ the *test algebra* of $A$ and say that *A has tests*. We denote the class of t-semirings by TS. A t-semiring is a *Kt-semiring* or *Kleene algebra with tests* if the t-semiring is also a K-semiring [Kozen 1997]. The class of Kleene algebras with test is denoted by KAT.

We will henceforth use letters $a, b, c, \ldots$ for arbitrary semiring elements (actions) and the letters $p, q, r, \ldots$ for tests (propositions). Moreover, we denote by $p'$ the complement of test $p$ in $\mathsf{test}(A)$ and by $p \sqcap q$ the meet of $p$ and $q$.

LEMMA 3.5. IS $\subseteq$ TS.

PROOF. Let $A \in \mathsf{IS}$. If $0 = 1$ then the claim $A \in \mathsf{TS}$ is trivially satisfied. Otherwise, let $\mathsf{test}(A) = \{0, 1\}$ with $p \sqcup q = p + q$, $p \sqcap q = pq$ for all $p, q \in \mathsf{test}A$ and $1' = 0$, $0' = 1$. This yields a Boolean subalgebra. □ □

We call t-semirings with test algebra $\{0, 1\}$ *discrete*.

LEMMA 3.6. *Let* $p, q \in \mathsf{test}(A)$ *for some* $A \in \mathsf{TS}$.

(i)  $pp = p$.

(ii) $pq = p \sqcap q$.

PROOF. By Lemma 3.2, $pq \leq p \sqcap q$.

(i)  $p = p1 = p(p + p') = pp + pp' \leq pp + (p \sqcap p') = pp + 0 = pp \leq p1 = p$.

(ii) Similar to the first part of the proof of Lemma 3.4, using idempotence of tests.
    □

□

The following lemma collects some properties of TS which will be needed for computing with abstract image and preimage operations in Section 6.

LEMMA 3.7. *Let* $A \in \mathsf{TS}$ *with* $a \in A$ *and* $p, q \in \mathsf{test}(A)$.

(i)  *The following properties are equivalent.*

$$pa \leq aq,$$
$$aq' \leq p'a,$$
$$paq' \leq 0,$$
$$pa = paq.$$

(ii) *The following properties are equivalent.*

$$ap \leq qa$$
$$q'a \leq ap',$$
$$q'ap \leq 0,$$
$$ap = qap.$$

PROOF. We only show (i). The proofs of (ii) are symmetric.

(1) $pa \leq aq \Rightarrow aq' \leq p'a$.

$$aq' = 1aq' = (p + p')aq' = paq' + p'aq' \leq aqq' + p'a = a0 + p'a = p'a.$$

(2) $aq' \leq p'a \Rightarrow paq' \leq 0$.

$$paq' \leq pp'a = 0a = 0.$$

(3) $paq' \leq 0 \Rightarrow pa = paq$.

$$pa = pa1 = pa(q + q') = paq + paq' = paq.$$

(4) $pa = paq \Rightarrow pa \leq aq$.

$$pa = paq \leq aq.$$

□

□

### 3.3  Tests in b-Monoids

When $A$ is a b-monoid with uniquely defined complement $\overline{a}$ for each $a \in A$, the set of subidentities is much better behaved. Now complements in $\mathsf{sid}(A)$ can be defined as restrictions of complements in $A$, viz. as $p' = 1 \sqcap \overline{p}$. The properties $1' = 0$, $0' = 1$, $p + p' = 1$, $pp' = 0$ are easily verified. Using the restricted complement we can show that all subidentities are multiplicatively idempotent, since

$$p = 1p = (p + p')p = pp + p'p = pp + 0 = pp.$$

Consequently, $pq = p \sqcap q$ and the whole set $\mathsf{sid}(A)$ is a Boolean subalgebra of $A$ (cf. [Desharnais and Möller 2001]).

The following lemma is the key to our comparison of the domain operations in t-semirings and b-monoids in Subsection 4.6.

LEMMA 3.8.  *(i)  Let $A$ be a b-monoid. Let $a \in A$ and $p \in \mathsf{sid}(A)$. Then*

$$a \leq p\top \Leftrightarrow a \leq pa. \tag{19}$$

*(ii)  There is a d-monoid $A$ such that the implication*

$$a \leq p\top \Rightarrow a \leq pa$$

*does not hold.*

PROOF. (i)  $a \leq pa$ implies $a \leq p\top$, since $a \leq \top$.

We now show that $a \leq p\top$ implies $a \leq pa$. By lattice algebra, $a \leq p\top$ iff $a = a \sqcap p\top$. We calculate

$$\begin{aligned}
a &= a \sqcap p\top \\
&= a \sqcap p(a + \overline{a}) \\
&= (a \sqcap pa) + (a \sqcap p\overline{a}) \\
&= pa + (a \sqcap p\overline{a}) \\
&\leq pa + (a \sqcap \overline{a}) \\
&= pa.
\end{aligned}$$

(ii)  The i-semiring $A_4^1$ of Example 2.6 is clearly also a d-monoid with $\top = b$, since the natural ordering is a chain. It satisfies $a = ab = a\top$, but $a \not\leq 0 = aa$.  □

□

### 3.4  Example Structures

We now consider some models of $\mathsf{TS}$ and $\mathsf{KAT}$. First, note that all examples by Conway from Section 2 (that is, Example 2.2 to Example 2.6) are discrete and therefore not very interesting.

EXAMPLE 3.9.  *In $\mathsf{REL}(A)$, there are $2^{|A|}$ subrelations of $\Delta$. They form a Boolean algebra with $P \sqcap Q = P \circ Q$ and $P' = \Delta - P$. For finite relations, in particular, this can be verified in the matrix representation.*  □

EXAMPLE 3.10.  *In $\mathsf{LAN}(\Sigma)$, the only subidentities are $\emptyset$ and $\{\varepsilon\}$. They also form the only possible test algebra; hence $\mathsf{LAN}(\Sigma)$ is always discrete. The example easily generalizes to arbitrary monoids.*  □

EXAMPLE 3.11. *In the tropical semiring, all elements are subidentities. However, except for $0$ and $\infty$, they are not idempotent. Thus the only possible test algebra consists of the elements $0$ and $\infty$.* □

EXAMPLE 3.12. *In the max-plus semiring, the only subidentities are $-\infty$ and $0$. These two elements also form the only possible test algebra.* □

EXAMPLE 3.13. *In the path i-semiring $\mathsf{PAT}(\Sigma)$ over $\Sigma$ (cf. Example 2.10), the subidenties $P \subseteq \Sigma \cup \{\varepsilon\}$ can be considered as modelling sets of nodes or states, where $\varepsilon$ also serves as the only "pseudo-node" or "pseudo-state" in an empty sequence.* □

## 4. DOMAIN

In this section, we introduce several equivalent axiomatizations of the domain operation on TS, among them a purely equational one. For a differentiated picture, we present two notions of different expressive power:

—A notion of *predomain* that suffices for deriving many natural properties of domain, as we will show in Section 4.4.
—A notion of *domain* that is important for more advanced applications, notably for treating modal operators.

We also show independence of the respective axioms, discuss extensions to b-monoids, quantales and relation algebras, and provide examples for the standard models.

### 4.1 Domain in the Relational i-Semiring

In order to motivate our abstract definitions, consider again the relational i-semiring of Example 2.7. Let $R \subseteq A \times A$ for some set $A$. Then the domain of $R$ is given by the set

$$\{a \in A \mid \exists\, b \in A \,.\, (a,b) \in R\}.$$

For our abstraction to t-semirings, it should be represented as a binary relation instead, viz. as the subidentity

$$\delta(R) = \{(a,a) \in A \times A \mid \exists b \in A \,.\, (a,b) \in R\}.$$

In the following subsections, we will propose algebraic point-free characterizations of a predomain and a domain operation. We leave it to the reader to show that they are consistent with the relational semiring. But first, let us replace the set-theoretic characterization of domain by two more algebraic ones.

First, $\delta(R)$ is the least solution for $X$ of the inclusion

$$R \subseteq X \circ R.$$

Second, using Example 3.9, the complement $\delta(R)'$ of $\delta(R)$ in the Boolean lattice of subidentities of $\mathsf{REL}(A)$ — the set of all pairs below $\Delta$ that are not in $\delta(R)$ — is the greatest solution for $X$ of the inclusion

$$X \circ R \subseteq \emptyset$$

under the constraint $X \subseteq \Delta$. Without this restriction, the greatest solution is $\overline{V \circ R^{\smile}}$, where $V$ denotes the universal relation and $R^{\smile}$ is the converse of $R$.

Since $\mathsf{REL}(A)$ is a complete lattice with respect to set-inclusion, both solutions are unique and the functional notation $\delta(R)$ is justified. In particular, $\delta(R) \subseteq \Delta$ is an immediate consequence of the definition in terms of a least solution. Since, according to Example 3.9, the subidentities of $\mathsf{REL}(A)$ form a Boolean algebra, Lemma 3.7 shows that the two definitions in terms of least and greatest solutions are indeed equivalent.

### 4.2 Preservers and Annihilators

As a first step in abstracting to semirings, we introduce some auxiliary concepts. Let $A \in \mathsf{IS}$ and $a, b \in A$. We say that $b$ *left-preserves* $a$ if $a \leq ba$, and that $a$ is *left-stable* under $b$ if $ba \leq a$. If $a = ba$ we say that $a$ is *left-invariant* under $b$. The concepts of *right-preservation*, *right-stability* and *right-invariance* are defined in a similar way. We say that $a$ is a *left annihilator* of $b$ if $ab = 0$, and a *right annihilator* if $ba = 0$. These concepts are useful in particular when $b \in \mathsf{sid}(A)$. Note that every element of $A$ is left- and right-invariant under 1 and that 0 is a left and right annihilator of every element of $A$.

We now use these concepts for abstracting the characterizations of domain of the previous subsection from the relational semiring to arbitrary idempotent semirings and test semirings.

LEMMA 4.1. *Let $A \in \mathsf{IS}$ and $a \in A$. The element $c \in A$ is the least left-preserver of $a$ iff*

$$\forall b \in A \,.\, c \leq b \Leftrightarrow a \leq ba. \tag{llp}$$

PROOF. We show that (llp) is equivalent to

$$a \leq ca, \tag{20}$$
$$a \leq ba \Rightarrow c \leq b. \tag{21}$$

Equation (21) is one direction of (llp). Setting $b = c$ in (llp) yields (20). Moreover, $a \leq ca \leq ba$ follows immediately from (20) and $c \leq b$. □ □

LEMMA 4.2. *In $\mathsf{IS}$,*

(i) *least left preservers are subidentities,*

(ii) *least left preservers are multiplicatively idempotent,*

(iii) *the set of least left preservers is a bounded distributive lattice with least element 0, greatest element 1, addition as join and multiplication as meet operation.*

PROOF. (i) Set $b = 1$ in (llp).

(ii) $cc \leq c$, follows from (i).
We have already seen in the proof of Lemma 4.1 that (llp) implies $a \leq ca$ (which is (20)), hence $a \leq cca$. Insertion into the right-hand side of (llp) yields $c \leq cc$.

(iii) This follows from (i) and (ii) with Lemma 3.4. □

□

An analogous treatment of greatest left annihilators in IS is, however, not straightforward. By Lemma 3.3, subidentities are in general not multiplicatively idempotent; meets and therefore complements need not in general exist.

There are two obvious solutions:

(1) Greatest left annihilators can be defined in terms of least left preservers if the distributive lattice of least left preservers can be extended to a Boolean lattice and if the search for a greatest solution is restricted to this Boolean lattice. This extension is possible by the representation theorem for distributive lattices (cf. [Birkhoff 1984]) according to which every distributive lattice can be isomorphically embedded into some field of sets.
(2) The considerations can be specialized from IS to TS. Then, least left preservers and greatest left annihilators can be defined as mappings into the set of tests.

Here, we choose the second alternative because of its simplicity, naturalness and technical convenience. Then, in particular, least left preservers and greatest left annihilators are multiplicatively idempotent subidentities by definition. Since domain elements are essentially abstractions of sets, they should possess a Boolean structure.

For the remainder, we will therefore restrict our attention to test semirings.

LEMMA 4.3. *Let $A \in \mathsf{TS}$ and $a \in A$. Then $c$ is the greatest left-annihilator of $a$ iff*

$$\forall p \in \mathsf{test}(A) \, . \, p \leq c \Leftrightarrow pa \leq 0. \tag{gla}$$

PROOF. We must show that (gla) is equivalent to

$$ca \leq 0, \tag{22}$$
$$pa \leq 0 \Rightarrow p \leq c. \tag{23}$$

The calculations are similar to those in the proof of Lemma 4.1.     □

It follows from properties of the partial ordering that least left-preservers and greatest left-annihilators are unique if they exist.

If the test algebra is complete then indeed they always exist, since property (gla) is easily seen to be closed under suprema.

The following lemma shows the relation between the least left-preserver and the greatest left-annihilator in a test semiring.

PROPOSITION 4.4. *Let $A \in \mathsf{TS}$. For all $a \in A$, let $c$ be the least left-preserver of $a$ in $A$ and let $g$ be the greatest left-annihilator of $a$ in $A$. Then $c = g'$.*

PROOF. Using Lemma 3.7, (llp) and (gla), we calculate

$$c \leq p \Leftrightarrow a \leq pa \Leftrightarrow p'a \leq 0 \Leftrightarrow p' \leq g \Leftrightarrow g' \leq p.$$

Thus $c = g'$ by the principle of indirect inequality.     □

### 4.3  Defining Predomain

We now define a predomain operation on test semirings using least left preservers. Proposition 4.4 provides an equivalent characterization in terms of greatest left annihilators. Moreover, we provide a further equivalent characterization in terms of two simple equations. We also show independence of the equational axioms and that predomain exists and is uniquely defined for each test semiring.

*Definition* 4.5. A structure $(A, \delta)$ is a *t-semiring with predomain* (a *$\delta$-semiring*) if $A \in \mathsf{TS}$ and the *predomain operation* $\delta : A \to \mathsf{test}(A)$ satisfies (llp), that is, for all $a \in A$ and $p \in \mathsf{test}(A)$,

$$\delta(a) \leq p \Leftrightarrow a \leq pa. \tag{llp}$$

The class of t-semirings with predomain is denoted $\mathsf{TSP}$.

The predomain is always unique if it exists, since least elements in a partial order are always unique.

We distinguish between predomain and domain, since, as already noted, the weaker definition suffices for deriving many natural properties.

PROPOSITION 4.6. $\mathsf{TSP}$ *is precisely the class of* $\mathsf{TS}$ *where each* $A \in \mathsf{TSP}$ *is enriched by a mapping* $\delta : A \to \mathsf{test}(A)$ *that satisfies, for all* $a \in A$ *and* $p \in \mathsf{test}(A)$,

$$\delta(a) \leq p \Leftrightarrow p'a \leq 0. \tag{gla}$$

PROOF. Immediate from Proposition 4.4. □ □

Note that here we have used the lattice-theoretic dual of the greatest left annihilator property.

We now present an equational characterization of predomain.

THEOREM 4.7. $\mathsf{TSP}$ *is precisely the class of* $\mathsf{TS}$ *where each* $A \in \mathsf{TSP}$ *is enriched by a mapping* $\delta : A \to \mathsf{test}(A)$ *that satisfies, for all* $a \in A$ *and* $p \in \mathsf{test}(A)$, *the two equations*

$$a \leq \delta(a)a, \tag{d1}$$
$$\delta(pa) \leq p. \tag{d2}$$

PROOF. We prove a somewhat stronger statement. First, we show that (d1) is equivalent to

$$\delta(a) \leq p \Rightarrow a \leq pa, \tag{24}$$

which is one direction of (llp). Obviously, (24) implies (d1), setting $p = \delta(a)$. For the converse direction, $a \leq \delta(a)a$ and $\delta(a) \leq p$ imply $a \leq pa$ by monotonicity of multiplication.

Second, we show that (d2) is equivalent to

$$a \leq pa \Rightarrow \delta(a) \leq p, \tag{25}$$

which is the other direction of (llp). Obviously, (25) implies (d2), instantiating $a$ by $pa$ and using multiplicative idempotence of $p$. For the converse direction, observe that $a \leq pa$ implies $a = pa$, since $p \leq 1$. Thus $\delta(a) = \delta(pa) \leq p$ by (d2). □

COROLLARY 4.8. $\mathsf{TSP}$ *is a variety.*

We have thus presented three equivalent axiomatizations for predomain. They are all of particular interest. The use of the equivalences (llp) and (gla) allows us to reduce certain $\mathsf{TSP}$-expressions to $\mathsf{TS}$-expressions that do not mention domain.

Moreover, both capture the basic algebraic intuition behind this concept. The two equational axioms (d1) and (d2) are perhaps less intuitive, but very beneficial for several reasons. First, they allow us to classify t-semirings with domain in Section 8. Second, they enable us to connect $\delta$-semirings with modal algebras and logics, which is, however, beyond the scope of this work. Third, they support a simple check whether some given mapping in some test semiring is a domain operation. The three axiomatizations taken together give us maximal flexibility in calculations.

We now show that the equational axiomatization is minimal:

THEOREM 4.9. *(d1) and (d2) are independent in* TS.

PROOF. We provide t-semirings in which precisely one of these axioms holds.

Set $\delta(0) = \delta(1) = 1$ in the Boolean semiring $A_2$ (Example 2.2). Then (d1) holds by neutrality of 1. But $\delta(01) = 1 \not\leq 0$. Thus (d2) does not hold.

Set $\delta(0) = \delta(1) = 0$ in the same (and only) Boolean semiring. Then (d2) holds by leastness of 0. But $1 \not\leq 0 = 01 = \delta(1)1$. Thus (d1) does not hold. □ □

We will see in the following subsection that (d1) and (d2) together imply that $\delta(a) = 0$ iff $a = 0$.

We now show that there also is always a meaningful — even if not very interesting — predomain definition for an i-semiring by choosing the discrete algebra of tests.

LEMMA 4.10. *A discrete t-semiring admits precisely one predomain operation.*

PROOF. Let $A \in$ TS. The mapping $f$ defined by $f : 0 \mapsto 0$ and $f : a \mapsto 1$ for all $0 \neq a \in A$ satisfies (d1) and (d2).

For (d1), if $\delta(a) = 0$ then $a = 0$. Hence $\delta(a)a = \delta(0)0 = 0 = a$. Otherwise, if $a \neq 0$ then $\delta(a) = 1$. Hence $\delta(a)a = 1a = a$.

For (d2), if $\delta(pa) = 0$ then (d2) holds trivially. Otherwise, if $\delta(pa) = 1$ then $pa \neq 0$ and therefore also $p \neq 0$. Thus $p = 1$ and (d2) also holds.

Thus $\delta$ is a well-defined predomain operation for $A$.

Finally, uniqueness is immediate from Lemma 4.11 (i), which will be shown in the next subsection. □ □

Let us conclude this section with a general remark. In opposition to relational semirings, the elements of general test semirings are intensional, that is, they are not completely determined by the elements of the associated test algebra. For a given i-semiring there may be many test algebras that can be embedded. These and the choice of the associated (pre)domain operation determine the precision of measuring properties of the Kleenean elements. Thus our definition of domain leaves the possibility of distinguishing not only between extensional and intensional behavior, but also between different degrees of intensionality.

### 4.4 Predomain Calculus

A look at the relational semiring shows that the domain operation has further useful and interesting algebraic properties. We will now show that many of them can already be derived from our simple definition of predomain. We will see, however, in the remaining sections that an additional equational axiom is needed for more advanced applications. The statements of this section are useful for a more intuitive understanding of domain. They also serve as the basic library of rules in a domain calculus.

Here, we list algebraic properties of domain without discussing their counterparts in the relational model. We leave this exercise to the reader or appeal to intuition.

LEMMA 4.11. *Let $A \in \mathsf{TSP}$. Let $a, b \in A$, $p \in \mathsf{test}(A)$ and $q \in \mathsf{sid}(A)$.*

(i)    *$\delta$ is fully strict:*
$$\delta(a) \leq 0 \Leftrightarrow a \leq 0. \tag{26}$$

(ii)    *$\delta$ is additive:*
$$\delta(a + b) = \delta(a) + \delta(b). \tag{27}$$

(iii)    *$\delta$ is monotonic:*
$$a \leq b \Rightarrow \delta(a) \leq \delta(b). \tag{28}$$

(iv)    *$\delta$ is an identity on tests:*
$$\delta(p) = p. \tag{29}$$

(v)    *$\delta$ is idempotent:*
$$\delta(\delta(a)) = \delta(a). \tag{30}$$

(vi)    *$\delta$ yields a left invariant:*
$$a = \delta(a)a. \tag{31}$$

(vii)    *$\delta$ satisfies an import/export law:*
$$\delta(pa) = p\delta(a). \tag{32}$$

(viii)    *$\delta$ satisfies a decomposition law:*
$$\delta(ab) \leq \delta(a\delta(b)). \tag{33}$$

(ix)    *$\delta$ commutes with the complement operation on tests:*
$$\delta(p)' = \delta(p'). \tag{34}$$

PROOF. (i)    $\delta(a) \leq 0 \Leftrightarrow a \leq 0a \Leftrightarrow a \leq 0$ follows from (llp).

(ii)    Using (gla), we calculate
$$\begin{aligned}\delta(a+b) \leq p &\Leftrightarrow p'(a+b) \leq 0 \\ &\Leftrightarrow p'a + p'b \leq 0 \\ &\Leftrightarrow p'a \leq 0 \wedge p'b \leq 0 \\ &\Leftrightarrow \delta(a) \leq p \wedge \delta(b) \leq p \\ &\Leftrightarrow \delta(a) + \delta(b) \leq p.\end{aligned}$$

The claim then follows from the principle of indirect inequality.

(iii)    Using (27), this is a standard result from lattice theory.

(iv)    $p \leq \delta(p)p \leq \delta(p)$ follows immediately from (d1) and $p \leq 1$. $\delta(p) = \delta(p1) \leq p$ follows immediately from (d2).

(v)     Immediate from (iv).

(vi)    By (d1) it remains to show that $\delta(a)a \leq a$, which is evident, since $\delta(a) \in \mathsf{test}(A)$.

(vii)   By Boolean algebra and (27) we have $\delta(a) = \delta(pa) + \delta(p'a)$. Now

$$p\delta(a) = p\delta(pa) + p\delta(p'a) = \delta(pa),$$

since $\delta(pa) \leq p$ and $\delta(p'a) \leq p'$ by (d2).

(viii)  By (llp) it suffices to show that $ab \leq \delta(a\delta(b))ab$. We calculate

$$ab \leq a\delta(b)b \leq \delta(a\delta(b))a\delta(b)b \leq \delta(a\delta(b))ab.$$

(ix)    Immediate from (iv). □

□

Most of these equations are also useful for simplifying terms on t-semirings that mention domain.

To conclude this section we note that in presence of a greatest element the predomain operation takes part in a proper Galois connection.

LEMMA 4.12. *If $A \in \mathsf{TSP}$ has a greatest element $\top$ then for all $a \in A$ and $p \in \mathsf{test}(A)$*

$$\delta(a) \leq p \Leftrightarrow a \leq p\top.$$

PROOF. ($\Rightarrow$) $a = \delta(a)a \leq pa \leq p\top$.
($\Leftarrow$) $\delta(a) \leq \delta(p\top) = p$. □

Note, however, that this Galois connection cannot be used as an alternative definition of domain, since it does not give all the properties we have derived so far. This works only in the case of b-monoids (see Section 4.6).

### 4.5 Locality and Domain Definition

Our definition of domain for t-semirings is not yet complete. There is a natural property of domain — called *locality* — that holds in the relational model but which is independent of (d1) and (d2). Namely

$$\delta(R \circ S) = \delta(R \circ \delta(S))$$

holds for all $R, S \in A \times A$, where $A$ is a set. We leave the verification to the reader. Intuitively, for computing the domain of a relation $R \circ S$, information about the domain of $S$ suffices; information about the inner structure or the codomain of $S$ is not needed.

In TSP, only one half of locality is derivable, as Lemma 4.11 (viii) shows, the other half is independent.

LEMMA 4.13. *There is an $A \in \mathsf{TSP}$ in which*

$$\delta(a\delta(b)) \leq \delta(ab)$$

*does not hold for all $a, b \in A$.*

PROOF. Consider again the discrete t-semiring $A_3^2$ of Example 2.4. According to Lemma 4.10, the mapping $f : 0 \mapsto 0$, $f : 1 \mapsto 1$, and $f : a \mapsto 1$ is a predomain operation. Then $f(af(a)) = f(a1) = 1$ and $f(aa) = f(0) = 0$. That is, $f(aa) \leq f(af(a))$ holds, but not $f(aa) = f(af(a))$. □ □

Due to independence of locality, we add the property of Lemma 4.13 to the predomain axioms to define the domain operation. However, we would like to distinguish between the two definitions, since in many applications, that property is not needed.

*Definition* 4.14. A *t-semiring with domain* (a $\hat{\delta}$-*semiring*) is a $\delta$-semiring in which the predomain operation $\hat{\delta} : A \to \mathsf{test}(A)$ also satisfies

$$\hat{\delta}(a\hat{\delta}(b)) \leq \hat{\delta}(ab), \tag{dloc}$$

for all $a, b \in A$. We denote the class of t-semirings with domain by $\mathsf{TSD}$.

We also use the term $\delta$-*locality* to distinguish locality of domain from that of codomain.

We now impose a necessary and sufficient condition such that a discrete $\delta$-semiring is also a $\hat{\delta}$-semiring. In analogy to the definition of an integral domain in ring theory, we say that a semiring $A$ is *integral* if it has no zero divisors, that is,

$$ab \leq 0 \Rightarrow a \leq 0 \vee b \leq 0. \tag{35}$$

holds for all $a, b \in A$.

LEMMA 4.15. *A discrete t-semiring is a $\hat{\delta}$-semiring iff it is integral.*

PROOF. Let $A$ be a discrete t-semiring. From Lemma 4.10 we know that $f$ defined by $f : 0 \mapsto 0$ and $f : a \to 1$ for all $0 \neq a \in A$ is the unique predomain operation on $A$. Thus $A$ is a $\delta$-semiring.

Let $A$ be integral. We must show that $f(af(b)) \leq 0$ whenever $f(ab) \leq 0$. So let $f(ab) \leq 0$. Then the construction of $f$ implies that $ab \leq 0$, hence $a \leq 0$ or $b \leq 0$, since there are no zero divisors. In the first case,

$$f(af(b)) = f(0f(b)) = f(0) = 0,$$

by construction of $f$. In the second case

$$f(af(b)) = f(af(0)) = f(a0) = f(0) = 0,$$

again by construction of $f$.

Now assume that $f$ satisfies (dloc), that is, $f(af(b)) \leq f(ab)$, and let $ab \leq 0$. Thus $f(af(b)) \leq f(ab) \leq 0$ and hence $af(b) \leq 0$ by construction of $f$. There are two cases.

If $f(b) = 1$ then $af(b) = a1 = a$. Hence $af(b) \leq 0$ implies $a \leq 0$.

If $f(b) = 0$ then $b = 0$ by construction of $f$.

Thus $ab \leq 0$ implies $a \leq 0$ or $b \leq 0$, that is, $A$ is integral. □ □

We will now show that this condition can be generalized to a sufficient condition on non-discrete t-semirings.

LEMMA 4.16. *Every integral $\delta$-semiring is a $\hat{\delta}$-semiring.*

PROOF. Let $A$ be integral. Thus $ab \leq 0$ implies $a \leq 0$ or $b \leq 0$. We use the principle of indirect inequality and show

$$\delta(ab) \leq p \Rightarrow \delta(a\delta(b)) \leq p.$$

Using Proposition 4.6,

$$\begin{aligned} \delta(ab) \leq p &\Leftrightarrow p'ab \leq 0 \\ &\Rightarrow p'a \leq 0 \vee b \leq 0 \\ &\Leftrightarrow \delta(a) \leq p \vee \delta(b) \leq 0. \end{aligned}$$

In the first case $\delta(a\delta(b)) \leq \delta(a) \leq p$. In the second case,

$$\delta(a\delta(b)) = \delta(a0) = \delta(0) = 0 \leq p.$$

□ □

### 4.6 Domain in b-Monoids

Definitions for predomain have originally been given for b-monoids and b-quantales (cf. [Aarts 1992; Möller 1999; Desharnais and Möller 2001]). There, the situation is considerably simpler.

First, as we have pointed out in Section 3, the entire set of subidentities of a b-monoid forms a Boolean sublattice and therefore a suitable test algebra. Second, predomain can now be *defined* in terms of the Galois connection (cf. Lemma 4.12)

$$\delta(a) \leq p \Leftrightarrow a \leq p\top, \tag{36}$$

from which the equational axioms

$$a \leq \delta(a)\top, \tag{37}$$
$$\delta(p\top) \leq p \tag{38}$$

are obtained in a generic way. Note that (36) and the equations (37) and (38) are just (llp) and the equations (d1) and (d2), when $\top$ is replaced by $a$.

In fact, by Lemma 3.8, in b-monoids, (37) is equivalent to (d1) and (38) is equivalent to (d2). However, (llp) does not express a Galois connection. It is therefore rather surprising that (d1), (d2) are equivalent to (llp) in TSP. Moreover, standard Galois theory would suggest that also (28) is needed as an equational axiom. The fact that this is not the case in TSP and therefore also in b-monoids is rather surprising.

We will now show formally that for b-monoids, the definition of predomain via least left preservers and that via the Galois connection coincide. We also show that the requirement on the monoid cannot be much relaxed. This means that our definition of pre-domain is really non-trivial.

LEMMA 4.17.

(i) For every b-monoid, (llp) and (36) are equivalent.
(ii) There is a d-monoid in which (llp) holds, but not (36).

PROOF. (i) Immediate from Lemma 3.8 (i).

(ii) Let $\delta(0) = 0$ and $\delta(a) = 1$ for all $a \neq 0$ in the discrete t-semiring $A_4^1$ (Example 2.6). Then clearly (llp) holds, but (36) does not hold by Lemma 3.8 (ii). □ □

In a b-quantale, domain is a priori well defined by the Galois connection.

### 4.7 Example Structures

We now consider some models in TSP and TSD.

EXAMPLE 4.18. *In the Boolean semiring $A_2$ (Example 2.2), the test algebra coincides with $A_2$. Setting $\delta(x) = 0 \Leftrightarrow x = 0$ is compatible with the definition of $f$ in Lemma 4.10. Thus it satisfies (d1) and (d2). Since $A_2$ is integral, also (dloc) holds. Moreover, this definition is unique.* □

EXAMPLE 4.19. *In $A_3^2$ (Example 2.4), the test algebra is $\{0,1\}$. Setting $\delta(0) = 0$, $\delta(a) = 1$ and $\delta(1) = 1$ is compatible with the definition of $f$ in Lemma 4.10. Thus it satisfies (d1) and (d2). Since $A_3^2$ is integral, also (dloc) holds. Moreover, this definition is unique.* □

EXAMPLE 4.20. *The only possible test algebra of the language i-semiring (Example 2.9) is $\{\emptyset, \{\varepsilon\}\}$. We set $\delta(\emptyset) = \emptyset$ and $\delta(L) = \{\varepsilon\}$ for all $\emptyset \neq L \subseteq \Sigma^*$. This is compatible with the definition of $f$ in Lemma 4.10. Thus it satisfies (d1) and (d2). Since the language model is integral (since it is free), also (dloc) holds. Moreover, this definition is unique. The example easily generalizes to arbitrary monoids.* □

EXAMPLE 4.21. *In the path i-semiring (Example 2.10), the test algebra is $2^{\Sigma \cup \{\varepsilon\}}$. For $S \subseteq \Sigma^*$, the set $\delta(S)$ consists of all starting (pseudo-)nodes/states of sequences in $S$. Although the semiring is not integral, (dloc) holds.* □

EXAMPLE 4.22. *In the tropical semiring, the test algebra consists solely of $0$ and $\infty$. Taking $\delta(\infty) = \infty$ and $\delta(n) = 0$ is compatible with the definition of $f$ in Lemma 4.10. Thus it satisfies (d1) and (d2). Since the tropical semiring is integral, also (dloc) holds. Moreover, this definition is unique.* □

These examples show that our definition of domain is meaningful in all the usual models, although non-trivial only in the relational model and the path model.

### 5. CODOMAIN

In this section, we introduce an equational axiomatization of codomain for idempotent semirings and two concepts of *duality*, one based on the *opposite* of a semiring, the other one based on the operation of *converse*, that allow an automatic transfer between statements about domain and those about codomain and save half of the work in proofs.

The definition of *codomain* parallels that of *domain*. For a set-theoretic relation $R \subseteq A \times A$, it is defined as

$$\rho(R) = \{b \in A \mid \exists a \in A \,.\, (a,b) \in R\}.$$

For a t-semiring this suggests to define a codomain operation as a least right preserver or a greatest right annihilator. Similarly to domain, there is a property of $\rho$-locality that is independent from the other postulates.

### 5.1 Codomain Definition

As usual, the *opposite* of a semiring $(A, +, \cdot, 0, 1)$ is the structure $(A, +, \breve{\cdot}, 0, 1)$ where $a \breve{\cdot} b = b \cdot a$. We denote the opposite of a semiring $A$ by $A^{op}$.

*Definition* 5.1. (i) A *t-semiring with precodomain* (a *ρ-semiring*) is a structure $(A, \rho)$ such that $(A^{op}, \rho)$ is a semiring with predomain.

(ii) A *t-semiring with codomain* (a *$\hat{\rho}$-semiring*) is a structure $(A, \hat{\rho})$ such that $(A^{op}, \hat{\rho})$ is a semiring with domain.

LEMMA 5.2. *Let $A$ be a ρ-semiring.*

(i) *The mapping $\rho$ has type $A \to \text{test}(A)$.*

(ii) *For all $a \in A$, the element $\rho(a)$ is a least right preserver of $a$, that is, for all $p \in \text{test}(A)$,*

$$\rho(a) \leq p \Leftrightarrow a \leq ap. \tag{lrp}$$

(iii) *For all $a \in A$, the element $\rho(a)$ is a greatest right annihilator of $a$, that is, for all $p \in \text{test}(A)$,*

$$\rho(a) \leq p \Leftrightarrow ap' \leq 0. \tag{gra}$$

(iv) *$\rho$ satisfies the following two equations.*

$$a \leq a\rho(a), \tag{cd1}$$
$$\rho(ap) \leq p. \tag{cd2}$$

(iii) *$A$ is a $\hat{\rho}$-semiring if also the following equation holds.*

$$\hat{\rho}(\hat{\rho}(a)b) \leq \hat{\rho}(ab). \tag{cdloc}$$

The proof is immediate from the definition and the results for predomain and domain in Section 4. More generally, all results of that section carry over to precodomain and codomain. Therefore we will only quote properties of domain even when talking about the codomain operation.

We call a t-semiring with predomain and precodomain a *δρ-semiring* and a t-semiring with domain and codomain a *$\hat{\delta}\hat{\rho}$-semiring*. When we do not want to distinguish, we uniformly speak about *test semirings with domain* and denote the associated class by TSD.

LEMMA 5.3. *There is a non-integral $\hat{\delta}\hat{\rho}$-semiring.*

PROOF. We have seen that (d1), (d2), (dloc), and (cd1), (cd2), (cdloc), respectively, hold in the relational model. However it its obvious that set-theoretic relations need not be integral. Let $R$ relate all even numbers and $S$ all odd numbers on $\mathbb{N}$. Then $R \neq \emptyset \neq S$, but $RS = \emptyset$. □ □

The path algebra is another non-integral $\hat{\delta}\hat{\rho}$-semiring.

### 5.2 Codomain via Converse

In the relational semiring, it is evident that the domain of a relation is the codomain of its converse and vice versa. This coupling of domain and codomain via the concept of converse induces a second notion of symmetry or duality, besides the one based on opposition. As usual, the operation of converse in an i-semiring is required to be involutive, additive and contravariant.

*Definition* 5.4. (i) An i-semiring with *preconverse* is a structure $(A, ^\circ)$ such that $A$ is an i-semiring and $^\circ : A \to A$ is an operation that satisfies the following equations.

$$a^{\circ\circ} = a, \tag{c1}$$
$$(a+b)^\circ = a^\circ + b^\circ, \tag{c2}$$
$$(ab)^\circ = b^\circ a^\circ. \tag{c3}$$

(ii) An i-semiring with *weak converse* is an i-semiring with preconverse such that all $p \leq 1$ satisfy

$$p^\circ \leq p. \tag{c4}$$

(iii) An i-semiring with *converse* [Crvenkovič et al. 2000] is an i-semiring with preconverse that satisfies the equation

$$a \leq aa^\circ a. \tag{c5}$$

It is easy to show that the properties

$$1^\circ = 1, \tag{39}$$
$$0^\circ = 0, \tag{40}$$
$$a \leq b \Leftrightarrow a^\circ \leq b^\circ \tag{41}$$

hold in every i-semiring with preconverse. The equation

$$p^\circ = p \tag{42}$$

holds in every i-semiring with weak converse. Moreover, every i-semiring with converse is an i-semiring with weak converse.

Using the operation of converse we can express the duality between domain and codomain *within* the test semiring rather than at the meta-level.

PROPOSITION 5.5. *Let $A$ be a $\delta\rho$-semiring (or a $\hat\delta\hat\rho$-semiring) with weak converse. Then for all $a \in A$,*

$$\delta(a^\circ) = \rho(a), \tag{43}$$
$$\rho(a^\circ) = \delta(a). \tag{44}$$

PROOF. We only show (43), thus verifying that $\delta(a^\circ)$ satisfies Definition 5.1 of codomain.

(cd1) By (d1), $a^\circ \leq \delta(a^\circ)a^\circ$, thus

$$a = a^{\circ\circ} \leq (\delta(a^\circ)a^\circ)^\circ = a^{\circ\circ}(\delta(a^\circ))^\circ = a\delta(a^\circ).$$

(cd2) By (d2),

$$\delta((ap)^\circ) = \delta(p^\circ a^\circ) = \delta(pa^\circ) \leq p.$$

(cdloc) By (dloc),

$$\delta((ab)^\circ) = \delta(b^\circ a^\circ) = \delta(b^\circ \delta(a^\circ)) = \delta(b^\circ(\delta(a^\circ))^\circ) = \delta((\delta(a^\circ)b)^\circ).$$

The proof of (44) is dual. □ □

We could therefore take (43) as definition of codomain in a t-semiring with weak converse.

COROLLARY 5.6. *Let $A$ be a $\delta\rho$-semiring with weak converse. For all $a \in A$, $p \in \mathsf{test}(A)$,*

$$\delta(a^\circ p) = \rho(pa), \qquad (45)$$
$$\rho(a^\circ p) = \delta(pa). \qquad (46)$$

### 5.3 Equivalence of $\delta$-Locality and $\rho$-Locality

It may come as a surprise that domain and codomain enjoy perfect symmetry with respect to locality of composition. We prepare the proof by an auxiliary property.

LEMMA 5.7. *A $\delta\rho$-semiring $A$ satisfies (dloc) iff for all $a, b \in A$,*

$$ab \leq 0 \Leftrightarrow \rho(a)\delta(b) \leq 0. \qquad (47)$$

PROOF. We first show that (dloc) implies (47).

$$\begin{aligned}
ab \leq 0 &\Leftrightarrow \delta(ab) \leq 0 \\
&\Leftrightarrow \delta(a\delta(b)) \leq 0 \\
&\Leftrightarrow a\delta(b) \leq 0 \\
&\Leftrightarrow \rho(a) \leq \delta(b)' \\
&\Leftrightarrow \rho(a)\delta(b) \leq 0.
\end{aligned}$$

The first and third steps of the proof use (26), the second step uses (dloc), the fourth step uses (gra) and the last step is by Boolean algebra.

Now we show that (47) implies (dloc). First, by (32) $\rho(a)\delta(b) = \rho(a\delta(b))$ and therefore, by (26) and (47)

$$ab \leq 0 \Leftrightarrow a\delta(b) \leq 0. \qquad (48)$$

Using Boolean algebra, (48) thrice and Boolean algebra again we calculate

$$\begin{aligned}
\delta(ab) \leq p &\Leftrightarrow p'\delta(ab) \leq 0 \\
&\Leftrightarrow p'ab \leq 0 \\
&\Leftrightarrow p'a\delta(b) \leq 0 \\
&\Leftrightarrow p'\delta(a\delta(b)) \leq 0 \\
&\Leftrightarrow \delta(a\delta(b)) \leq p,
\end{aligned}$$

whence $\delta(ab) = \delta(a\delta(b))$ by the principle of indirect inequality. □ □

Since (47) is symmetric in $\delta$ and $\rho$, we obtain

COROLLARY 5.8. *A $\delta\rho$-semiring is a $\hat{\delta}$-semiring iff it is a $\hat{\rho}$-semiring.*

## 6. IMAGE AND PREIMAGE

In many applications, domain and codomain operations occur more specifically as *image* and *preimage* operations for some given test element. In the relational semiring, the preimage of a set $B \subseteq A$ under a relation $R \subseteq A \times A$ is defined as

$$R:B = \{x \in A \,|\, \exists y \in B \,.\, (x,y) \in R\}.$$

We leave it to the reader to verify that this is equivalent to the point-free definition $R:B = \delta(R \circ B)$. Dually, the image of $B$ under $R$ is defined as

$$B:R = \{y \in A \,|\, \exists x \in A \,.\, (x,y) \in R\},$$

which is equivalent to the point-free definition by $B:R = \rho(B \circ R)$.

As usual, we abstract this point-free definition from sets to semirings and define for every $A \in \mathsf{TSP}$ the *image* and the *preimage operator*, both denoted by $:$, as mappings of type $\mathsf{test}(A) \times A \to \mathsf{test}(A)$ and $A \times \mathsf{test}(A) \to \mathsf{test}(A)$ by

$$p:a = \rho(pa), \tag{49}$$
$$a:p = \delta(ap), \tag{50}$$

for all $a \in A$ and $p \in \mathsf{test}(A)$. We henceforth use this notation and avoid domain and codomain whenever this is appropriate. In particular, we often use $a:1$ and $1:a$ instead of $\delta(a)$ and $\rho(a)$. We also overload this notation to definitions with respect to $\hat{\delta}$ and $\hat{\rho}$. Since the preimage and the image operator are multiplications, we stipulate that they bind stronger than addition.

Moreover, since image and preimage are defined by codomain and domain and since codomain and domain are coupled via the concept of opposition, there is again an automatic transfer between properties of image and those of preimage. Like in previous sections, we therefore only mention properties of preimage and quote preimage properties even when talking about the image operation.

The following lemma connects preimage with least left preservation and annihilation. Like (llp) and (gla), this allows us to eliminate certain occurrences of preimage and image operators.

LEMMA 6.1. *Let $A \in \mathsf{TSP}$. For all $a \in A$ and $p \in \mathsf{test}(A)$,*

$$a:p \leq q \Leftrightarrow ap \leq qa, \tag{51}$$
$$a:p \leq q \Leftrightarrow q'ap \leq 0. \tag{52}$$

PROOF. Immediate from (llp), and Lemma 3.7, respectively. □ ◻

From (32) we get the following import/export rule for the preimage.

COROLLARY 6.2. *Let $A \in \mathsf{TSP}$. For all $a \in A$ and $p, q \in \mathsf{test}(A)$,*

$$p(a:q) = (pa):q. \tag{53}$$

Lemma 6.1 has the following consequence that couples preimage and image operations.

LEMMA 6.3. *Let $A \in \mathsf{TSP}$. The preimage and the image operation satisfy the following* exchange law. *For all $a \in A$ and $p \in \mathsf{test}(A)$,*

$$a:p \leq q \Leftrightarrow q':a \leq p'. \tag{54}$$

PROOF. Immediate from Lemma 6.1 and Lemma 3.7. □ ◻

The equivalence (54) is a weak analogue of the Schröder rule from the relational calculus. Lemma 6.3 has the following immediate consequence.

COROLLARY 6.4. *Let $A \in \mathsf{TSP}$. For all $a \in A$ and $p \in \mathsf{test}(A)$,*

$$(p:a)q \leq 0 \Leftrightarrow p(a:q) \leq 0. \tag{55}$$

The decomposition property becomes

$$p:(ab) \leq (p:a):b; \tag{56}$$

under (dloc) this becomes an equality.

Finally, locality yields the following interaction of domain with preimage and of codomain with image.

LEMMA 6.5. *Let $A \in \mathsf{TSD}$. Then for all $a, b \in A$,*

$$\hat{\delta}(ab) = a:\hat{\delta}(b), \tag{57}$$
$$\hat{\rho}(ab) = \hat{\rho}(a):b. \tag{58}$$

## 7. DOMAIN, CODOMAIN AND KLEENE STAR

So far, we have only investigated domain and codomain operations in test semirings, that is, in absence of the Kleene star operation. In fact, there is no need for further axioms in presence of the Kleene star. Therefore, in this section, we only need to investigate the interaction of domain and codomain and that of image and preimage with the Kleene star. It turns out that only image and preimage show nontrivial behaviour. In particular we will see that when the Kleene star is adapted to occur within domain and codomain operators, a finite equational axiomatization instead of the Horn clauses ($*$-3) and ($*$-4) is possible. Moreover, one of these equational axioms can be interpreted as an efficient reachability algorithm, when interpreted over finite relations; its proof is by a formal derivation from a less efficient specification.

Henceforth, K-semirings are called K$\delta$-semirings, K$\rho$-semirings, K$\hat{\delta}$-semirings, K$\hat{\rho}$-semirings, K$\delta\rho$-semirings and K$\hat{\delta}\hat{\rho}$-semirings, when they are extended by the respective operation(s) and defined by the respective axioms. Moreover, when we do not want to distinguish, we uniformly speak of *Kleene algebra with predomain* or *Kleene algebra with domain*. We denote the classes by $\mathsf{KAP}$ and $\mathsf{KAD}$.

First, the properties of the Kleene star from Lemma 2.1 have some trivial consequences for domain and codomain.

LEMMA 7.1. *Let $A \in \mathsf{KAP}$. Then for all $a \in A$,*

$$\delta(a)^* = 1, \tag{59}$$
$$\delta(a^*) = 1. \tag{60}$$

The Kleene star in combination with images or preimages has a much richer and more interesting behaviour. The following three statements show that preimages in combination with star satisfy expressions analogous to ($*$-1) and ($*$-2) for K-semirings. Like their counterparts in $\mathsf{KA}$, they are the working horses for many interesting derivations. We give variants for $\mathsf{KAD}$, because presence of (dloc), that is, $(ab):p = a:(b:p)$, allows a more compositional treatment of images and preimages.

LEMMA 7.2. *Let $A \in \mathsf{KAP}$. For all $a \in A$ and $p \in \mathsf{test}(A)$,*
$$p + a^* : (a : p) \leq a^* : p \geq p + a : (a^* : p).$$
*The inequalities become equations, when $A \in \mathsf{KAD}$.*

PROOF. By (9),
$$a^* : p = (1 + a^*a) : p = (1 : p) + (a^*a) : p \geq p + a^* : (a : p).$$
The last step uses (33). The second half of the claim is shown analogously. The equations follow by using (dloc) instead of (33). □ □

Note the analogy to (9) in $\mathsf{KA}$. By Lemma 7.2, $a^* : p$ is a fixed point of the mapping $\lambda x . p + a : x$.

LEMMA 7.3. *Let $A \in \mathsf{KAD}$. For all $a \in A$ and $p \in \mathsf{test}(A)$,*
$$a : p \leq p \Rightarrow a^* : p \leq p. \tag{61}$$

PROOF. Using Lemma 6.1 and (12), we calculate
$$a : p \leq p \Leftrightarrow ap \leq pa \Rightarrow a^*p \leq pa^* \Leftrightarrow a^* : p \leq p.$$
□ □

Lemma 7.3 can also be viewed as an assertion about invariants: an invariant of $a$ is also an invariant of $a^*$. Moreover, it has two important consequences. First, we will use it in the following lemma to derive variants of the statements of Lemma 7.2 that lead to more efficient evaluation of the expressions involved. Second, when the Kleene star is adapted to occur only within preimages, we will show in the following lemma that there are even equivalent equational characterizations.

LEMMA 7.4. *Let $A \in \mathsf{KAD}$. Let $a \in A$ and $p, q \in \mathsf{test}(A)$. The following properties are equivalent and hence by Lemma 7.3 hold in $\mathsf{KAD}$.*
$$a : p \leq p \Rightarrow a^* : p \leq p, \tag{61}$$
$$a : p + q \leq p \Rightarrow a^* : q \leq p, \tag{62}$$
$$a^* : p \leq p + a^* : (p'(a : p)), \tag{63}$$
$$a^* : p = p + (ap')^* : (a : p). \tag{64}$$

PROOF. We first show that (61), (62) and (63) are equivalent.

(61) implies (62). $a : p + q \leq p$ iff $a : p \leq p$ and $q \leq p$ and therefore $a^* : p \leq p$ by the assumption. Hence also $a^* : q \leq p$ by monotonicity.

(62) implies (63). For $a^* : p \leq p + a^* : (p'(a : p))$ it suffices by (62) to show that
$$p \leq p + a^* : (p'(a : p)),$$
$$a : (p + a^* : (p'(a : p))) \leq p + a^* : (p'(a : p)).$$
The first inequality is trivial. The second one is proved as follows.
$$\begin{aligned}
a : (p + a^* : (p'(a : p))) &= (a : p) + a : (a^* : (p'(a : p))) \\
&= (p + p')(a : p) + a : (a^* : (p'(a : p))) \\
&\leq p + p'(a : p) + a : (a^* : (p'(a : p))) \\
&= p + a^* : (p'(a : p)).
\end{aligned}$$

The third step uses $p(a\!:\!p) \leq p$; the last step uses Lemma 7.2.

(63) implies (61). Let $a^*\!:\!p \leq p + a^*\!:\!(p'(a\!:\!p))$ and assume $a\!:\!p \leq p$. Then

$$a^*\!:\!p \leq p + a^*\!:\!(p'(a\!:\!p)) \leq p + a^*\!:\!(p'p) = p + a^*\!:\!0 = p + 0 = p.$$

We now show that (62) implies (64) and that (64) implies (61). This yields simpler proofs than a direct circle.

(62) implies (64). First, $p + (ap')^*\!:\!(a\!:\!p) \leq p + a^*\!:\!(a\!:\!p) = a^*\!:\!p$ by monotonicity of the Kleene star, the fact that $p \leq 1$ and Lemma 7.2 with (dloc). For the converse direction, that is, $a^*\!:\!p \leq p + (ap')^*\!:\!(a\!:\!p)$, it suffices by (62) to show that

$$p \leq p + (ap')^*\!:\!(a\!:\!p),$$
$$a\!:\!(p + (ap')^*\!:\!(a\!:\!p)) \leq p + (ap')^*\!:\!(a\!:\!p).$$

The first inequality is trivial. The second one is proved as follows.

$$\begin{aligned}
a\!:\!(p + (ap')^*\!:\!(a\!:\!p)) &= a\!:\!p + (a(p+p'))\!:\!((ap')^*\!:\!(a\!:\!p)) \\
&= a\!:\!p + (ap)\!:\!((ap')^*\!:\!(a\!:\!p)) + (ap')\!:\!((ap')^*\!:\!(a\!:\!p)) \\
&\leq a\!:\!p + (ap)\!:\!1 + (ap)'\!:\!((ap')^*\!:\!(a\!:\!p)) \\
&= a\!:\!p + (ap')\!:\!((ap')^*\!:\!(a\!:\!p)) \\
&= (ap')^*\!:\!(a\!:\!p) \\
&\leq p + (ap')^*\!:\!(a\!:\!p).
\end{aligned}$$

The first two steps use additivity of domain, the third step uses $(ap)^*\!:\!(a\!:\!p) \leq 1$, the fourth step uses that $(ap)\!:\!1 = a\!:\!p$, the fifth step uses (61).

(64) implies (61). Assume $a\!:\!p \leq p$. Then

$$\begin{aligned}
a^*\!:\!p &= p + (ap')^*\!:\!(a\!:\!p) \\
&\leq p + (ap')^*\!:\!p \\
&= p + (ap')^*\!:\!((ap')\!:\!p) \\
&= p + (ap')^*\!:\!0 \\
&= p.
\end{aligned}$$

The third step uses Lemma 7.2, the fourth step uses that $(ap')\!:\!p = \delta(app') = \delta(0) = 0$, the fifth step uses (26). □

Note the analogy of (62) to $b + ac \leq c \Rightarrow a^*b \leq c$, that is, (∗-3).

COROLLARY 7.5. *Let $A \in \mathsf{KAD}$. For all $a, b, c \in A$ and $p \in \mathsf{test}(A)$,*

$$(ac)\!:\!p + b\!:\!q \leq c\!:\!p \Rightarrow (a^*b)\!:\!q \leq c\!:\!p. \tag{65}$$

PROOF. The claim follows immediately from (62), replacing $p$ by $c\!:\!p$, $q$ by $b\!:\!q$ and using (dloc).    □ □

Already Lemma 7.2 describes an unfolding step of the preimage operation. However, this is not the most efficient way of unfolding. In $a^*\!:\!p = p + a^*\!:\!(a\!:\!p)$, for instance, it is not necessary to perform a full $a$-iteration from $a\!:\!p$. Since all steps starting from $p$ have already been considered, it suffices to perform the $a$-iteration from $p'$-states. This is expressed by (64).

To conclude this section we give another example that shows that the image and preimage mappings in a KAD again induce a K-semiring.

EXAMPLE 7.6. *(An algebra of predicate transformers)* Let $A \in \mathsf{KAD}$ and consider for all $a \in A$ the set $F_A$ of mappings $f_a = \lambda x.(a:x)$. We write $f_a(p) = a:p$ and define addition and multiplication on $F_A$ by

$$(f_a \oplus f_b)(p) = f_a(p) + f_b(p), \tag{66}$$
$$(f_a \odot f_b)(p) = f_a(f_b(p)), \tag{67}$$

for all $p \in \mathsf{test}(A)$. Then it is easy to verify that $(F_A, \oplus, \odot, f_0, f_1)$ is a t-semiring with set of tests $\{f_p \mid p \in \mathsf{test}(A)\}$. Moreover, setting

$$f_a^*(p) = a^*:p, \tag{68}$$

we obtain

$$f_1 \oplus (f_a \odot f_a^*) = f_a^*, \tag{69}$$
$$f_1 \oplus (f_a^* \odot f_a) = f_a^* \tag{70}$$

from the first half of Lemma 7.2 and

$$f_b \oplus (f_a \odot f_c) \leq f_c \Rightarrow f_a^* \oplus f_b \leq f_c \tag{71}$$

from Corollary 7.5. Hence $(F_A, \oplus, \odot, f_0, f_1, (.)^*)$ is a left K-semiring. □

## 8. KLEENE ALGEBRAS AS VARIETIES

In this section we classify some of our results in the context of universal algebra.

We identify varieties with equational classes. A variety is *finitely based* if it can be axiomatized by a finite set of equations. The following lemma is immediate.

LEMMA 8.1. TSD *is a finitely based variety.*

The next lemma is not so immediate. It has been shown in [Kozen 1994b; Pratt 1990] that KA with a residuation operation is a finitely based variety. The same phenomenon might occur when adding a domain or codomain operation. The following lemma shows that this is not the case. A similar argument has been used in [Hollenberg 1997] for algebras related to PDL.

LEMMA 8.2. KAP *and* KAD *are not finitely based varieties.*

PROOF. In [Conway 1971], p. 106, Conway gives an algebra $A_p$ for showing that the algebra of regular events (cf. Example 2.9) is not finitely based. For every finite set of equations and every prime $p$ there is a particular valid equation parameterized by $p$ that is not deducible, and there is an algebra $A_p$ parameterized by $p$ that satisfies the finite set of equations, but not the given additional equation. According to Conway, every expression in the language of KA is equivalent to some sum of terms each of which is either 0 or 1 or is simultaneously 0-free, 1-free and +-free. This implies that in $A_p$, which is constructed from such normal form terms, $ab \leq 0$ implies that $a \leq 0$ or $b \leq 0$, thus the integral condition (35) holds.

Now, in presence of domain, we consider the discrete t-semiring on $A_p$. Then by Lemma 4.10 and Lemma 4.15, the mapping defined by $\delta(0) = 0$ and $\delta(a) = 1$ for all $0 \neq a \in A_p$ satisfies (d1), (d2) and (dloc). In particular $0 \neq 1$.

Thus the expansion of $A_p$ satisfies the finite set of equations and the domain axioms, but not the given additional equation. Consequently, the given finite set of equations is not complete for KAP and KAD. □ □

## 9. RECONSTRUCTING NOETHERICITY

In this section we demonstrate the expressive power and applicability of KAD in the field of termination analysis of programs. We show that concepts of Noethericity and well-foundedness can be algebraically reconstructed. We further show that our concepts subsume those in Cohen's $\omega$-algebra [Cohen 2000], an extension of KA with infinite iteration that is defined as a greatest fixed point by expressions similar to ($*$-1), ($*$-2), ($*$-3) and ($*$-4). Moreover, adapting a result by Goldblatt [Goldblatt 1985], we show that for transitive relations our concept is also equivalent to an algebraic variant of Löb's formula from modal logic [Bull and Segerberg 1984; Chellas 1980]. Finally, we show that some simple and well-known properties of well-founded relations can be calculated in KAD in a simple and elegant way.

Intuitively, a set-theoretic relation $R \subseteq A \times A$ is well-founded if there are no infinitely descending $R$-chains, that is, no infinite chains $x_0, x_1, \ldots$ such that $(x_{i+1}, x_i) \in R$. Moreover, $R$ is Noetherian if there are no infinitely ascending $R$-chains, that is, no infinite chains $x_0, x_1, \ldots$ such that $(x_i, x_{i+1}) \in R$.

Thus, $R$ is *not* well-founded if there is a non-empty set $P \subseteq A$ (denoting the infinite chain) such that for all $x \in P$ there exists some $y \in P$ with $(y, x) \in R$. This is equivalent to saying that $P$ is contained in the image of $P$ under $R$, that is,

$$P \subseteq P : R. \tag{72}$$

Consequently, if $R$ *is* well-founded, then only the empty set may satisfy (72).

### 9.1 Noethericity: Definition and Simple Properties

Abstracting to $A \in \mathsf{TSD}$, we say that $a$ is *well-founded* if for all $p \in \mathsf{test}(A)$,

$$p \leq p : a \Rightarrow p \leq 0. \tag{73}$$

Moreover, $a$ is *Noetherian* if for all $p \in \mathsf{test}(A)$,

$$p \leq a : p \Rightarrow p \leq 0. \tag{74}$$

We now calculate abstract algebraic variants of some simple and well-known properties of well-founded and Noetherian relations. Again, as in previous sections, we restrict our attention to Noethericity, which is expressed in terms of preimages. We do not explicitly mention well-foundedness properties that hold by duality in the opposite semiring. In the context of termination, reflexivity is not a desirable property, as we will see. Therefore the transitive closure $a^+ = aa^*$ is more interesting than $a^*$ itself. We say that $a$ is transitive, if $aa \leq a$.

Lemma 9.1. *Let $A \in \mathsf{KAD}$. Let $a, b \in A$ and let $0 \neq 1$.*

(i)  *0 is Noetherian.*
(ii)  *Every test $p \neq 0$ is not Noetherian.*
(iii)  *If $b$ is Noetherian and $a \leq b$, then $a$ is Noetherian.*

(iv)   If $a$ is Noetherian, then $a \sqcap 1 = 0$, that is, $a$ is irreflexive[2].

(v)   If $a \not\leq 0$ is Noetherian then $a \not\leq aa$, that is, $a$ is not dense.

(vi)   $a$ is Noetherian iff $a^+$ is Noetherian.

(vii)   $a^*$ is not Noetherian.

PROOF. (i)   Let $p \leq 0 : p$. Then $p \leq 0$, since $0 : p = 0$.

(ii)   Every such $p$ satisfies
$$p \leq p \cdot p = p : p.$$

(iii)   Let $a$ be Noetherian an let $b \leq a$. Then
$$p \leq b : p \Rightarrow p \leq a : p \Rightarrow p \leq 0.$$
Thus $b$ is Noetherian.

(iv)   We show that $0$ is the only common lower bound of $a$ and $1$. Indeed, let $p$ be such a lower bound. Then, since $p \leq a$, by (iii) $p$ is Noetherian. On the other hand, $p \leq 1$ and by (ii) we infer $p = 0$.

(v)   Let $a$ be dense and Noetherian. $a \leq aa$ implies $a : p \leq a : (a : p)$, by monotonicity. Thus $a : p \leq 0$ for all $p \in \mathsf{test}(A)$. The particular case $p = 1$ yields $a \leq 0$, a contradiction.

(vi)   Let $a$ be Noetherian and remember that $a^+ = aa^*$. We calculate
$$\begin{aligned} p \leq a^+ : p &\Rightarrow a^* : p \leq a^* : (a^+ : p) \\ &\Leftrightarrow a^* : p \leq a : (a^* : p) \\ &\Rightarrow a^* : p \leq 0 \\ &\Rightarrow 1 : p \leq 0 \\ &\Leftrightarrow p \leq 0. \end{aligned}$$
The second step uses (dloc), $a^* a^* = a^*$ and $aa^* = a^* a$. The third step uses Noethericity of $a$. The fourth step uses $1 \leq a^*$. Thus $a^+$ is Noetherian.
Now let $a^+$ be Noetherian. Then, by (iii) and $a \leq a^+$, $a$ is Noetherian.

(vii)   By (ii), $1$ is not Noetherian. Then $1 \leq a^*$ implies that $a^*$ is not Noetherian using (iii). □

### 9.2 Noethericity and $\omega$-Algebra

We now show how our definition of Noethericity is related to the one in Cohen's $\omega$-algebra. We do not introduce the axioms for this class. Intuitively, while an expression $a^*$ denotes finite non-deterministic iteration of $a$, $a^\omega$ denotes infinite iteration. As an $\omega$-regular expression, $a^\omega$ is intended to denote a set of words of infinite length or streams. Consequently, in $\omega$-algebra Noethericity of $a$ means absence of proper infinite iteration of $a$; thus $a^\omega = 0$. In our calculations we only need the following property.

$$a^\omega \leq aa^\omega. \tag{75}$$

---

[2]The proof will show that this particular meet exists.

LEMMA 9.2. *Let $A$ be an $\omega$-algebra that is also a $\delta$-semiring. Then for all $a \in A$, if $a$ is Noetherian then $a^\omega = 0$.*

PROOF. Let $a$ be Noetherian. Using (33) we obtain
$$\delta(a^\omega) \leq \delta(aa^\omega) \leq \delta(a\delta(a^\omega)) = a : \delta(a^\omega).$$
Thus $\delta(a^\omega) = 0$ by definition (74) of Noethericity. By Lemma 4.11(i) (strictness of domain), this is the case if and only if $a^\omega = 0$. □

The converse implication does not hold. In the language semiring we have $a^\omega = 0$ if $1 \sqcap a = 0$, but also
$$a \neq 0 \Rightarrow (\forall\, p\,.\, a : p\ = p).$$

Note that $\omega$-algebra can only express Noethericity, whereas TSD can express both Noethericity and well-foundedness.

### 9.3 Noethericity and the Löb Axiom

We now investigate an alternative characterization of Noethericity for transitive relations that is even equational. Remember that an element of a semiring is transitive if $aa \leq a$.

In modal logic, Noethericity of the underlying Kripke frame is characterized by Löb's axiom (cf. [Bull and Segerberg 1984; Chellas 1980])
$$\Box(\Box p \to p) \to \Box p.$$
For our purposes, the equivalent version $\Diamond p \to \Diamond (p \wedge \neg \Diamond p)$ is more convenient, since it can immediately be translated into KAD, using preimage operators resulting in
$$a : p \leq a : (p - a : p). \tag{76}$$
Here we have transcribed $\Diamond p$ into $a : p$, where $a$ is a Kleene element that represents the underlying Kripke frame, and $p - q$ stands for $pq'$.

We say that $a$ is *Löbian* if it satisfies (76). In the relational model, Löb's axiom states that $a$ is transitive and that there are no infinite $a$-chains. Note the similarity to (63).

We will now relate Löb's axiom and our notion of Noethericity. But first we need a technical lemma.

LEMMA 9.3. *Let $A \in$ KAD. Let $a \in A$ and $p, q \in \mathsf{test}(A)$.*

(i)   $a : p - a : q \leq a : (p - q)$,

(ii)  $a^+ : p = a : (p + a^+ : p)$.

PROOF. (i)   $a : p = a : (p(q + q')) = a : (pq) + a : (pq') \leq a : q + a : (pq')$. The result then follows from the definition of subtraction.

(ii)  Immediate from Lemma 7.2 and the definition of $a^+$. □

The following theorem is essentially due to Goldblatt [Goldblatt 1985].

THEOREM 9.4. *Let $A \in$ KAD and let $a \in A$.*

(i)   *$a$ is Noetherian if it is Löbian.*

(ii) If $a$ is Noetherian then for all $p \in \text{test}(A)$,
$$a:p \leq a^+ :(p - a:p). \tag{77}$$

(iii) $a$ is Löbian if it is Noetherian and transitive.

PROOF. (i)  Let $p \leq a:p$. Thus equivalently $p - a:p \leq 0$ by Boolean algebra. Using (76) we calculate
$$p \leq a:p \leq a:(p - a:p) \leq a:0 = 0.$$

(ii) First, observe that (77) is equivalent to $a:p - a^+ :(p - a:p) \leq 0$. Thus by Noethericity of $a$ it suffices to show that
$$a:p - a^+ :(p - a:p) \leq a:(a:p - a^+ :(p - a:p)).$$

We calculate
$$\begin{aligned}
a:p - a^+ :(p - a:p) &= a:p - a:((p - a:p) + a^+ :(p - a:p)) \\
&\leq a:(p - ((p - a:p) + a^+ :(p - a:p))) \\
&= a:((p - (p - a:p)) - a^+ :(p - a:p)) \\
&\leq a:(a:p - a^+ :(p - a:p)).
\end{aligned}$$

The first and second step use Lemma 9.3 (ii) and (i). The third step uses $p - (q + r) = (p - q) - r$, which holds in Boolean algebra. The fourth step uses $p - (p - q) = pq \leq q$, which holds again in Boolean algebra, and monotonicity.

(iii) For transitive $a$ we have $a = a^+$ as the following instantiation of ($*$-4) shows:
$$aa^* \leq a \Leftarrow a + aa \leq a.$$

Now the claim is immediate from (ii). □

The statement of Theorem 9.4 is closely related to the correspondence theory of modal logic. In this view, our property of Noethericity expresses a frame property, which is part of semantics, whereas our Löb axiom stands for a modal formula, which is part of syntax. In KAD we are able to express syntax and semantics in the same formalism. Moreover, while the traditional proof of the correspondence uses an (informal) semantic argument, our proof is entirely calculational. Further investigations of Noethericity in the context of KAD are outside the scope of the present paper.

## 10. RECONSTRUCTING HOARE LOGIC

In this section we consider another application of KAD: an algebraic representation of propositional Hoare logic. Establishing this kind of subsumption relation is a popular exercise for many logics and algebras for imperative programming languages. Hoare logic has, for instance, already been embedded into PDL [Fischer and Ladner 1979] and KAT [Kozen 2001]. Since KAD is an extension of KAT, our subsumption result is no surprise. However we believe that it is interesting for at least two reasons. First, in KAD, an encoding of the inference rules of the Hoare calculus is much more crisp and clear and so are their correctness proofs. Second, the properties of the standard partial correctness semantics [Loeckx and Sieber 1987;

Apt and Olderog 1997] for Hoare logic mirror precisely those of domain, so that KAD may be considered a natural abstract algebraic semantics for propositional Hoare logic. A point that takes KAD strictly beyond KAT in this context is the possibility to express the weakest precondition operator as

$$wlp(a,p) = (a:p')'.$$

However, to keep matters short, we stay with Hoare logic in this text and refer to [Möller and Struth 2003b] for a full account of *wlp* in KAD.

We start by encoding the relevant programming constructs in KA.

$$a\,;b = ab, \tag{78}$$
$$\text{if } p \text{ then } a \text{ else } b = pa + p'b, \tag{79}$$
$$\text{while } p \text{ do } a = (pa)^*p'. \tag{80}$$

We now briefly recall the syntax and semantics of Hoare logic. The basic formulas are *partial correctness assertions* of the form $\{p\}\ a\ \{q\}$, where $p$ and $q$ (the *precondition* and *postcondition*) denote Boolean expressions and $a$ denotes a program. Intuitively, $p$ models a property of the input states of a program, while $q$ models a property that is intended to hold at the output states. The program $a$ is intuitively interpreted as a relation between input and output states. Traditionally, the Hoare calculus uses the following inference rules for reasoning about programs.

Assignment $\quad\{p[e/x]\}\ x := e\ \{p\},$

Composition $\quad\dfrac{\{p\}\ a\ \{q\}\quad \{q\}\ b\ \{r\}}{\{p\}\ a\,;b\ \{r\}},$

Conditional $\quad\dfrac{\{p\wedge q\}\ a\ \{r\}\quad \{p'\wedge q\}\ b\ \{r\}}{\{q\}\ \text{if } p \text{ then } a \text{ else } b\ \{r\}},$

While $\quad\dfrac{\{p\wedge q\}\ a\ \{q\}}{\{q\}\ \text{while } p \text{ do } a\ \{p'\wedge q\}},$

Weakening $\quad\dfrac{p_1 \to p\quad \{p\}\ a\ \{q\}\quad q\to q_1}{\{p_1\}\ a\ \{q_1\}}.$

Assignment is a non-propositional inference rule that deals with the internal structure of states. It is therefore disregarded in this embedding. Following [Kozen 2001], we call the fragment of Hoare logic without assignments *propositional Hoare logic* (PHL). Following [Kozen 2001] further, we define partial correctness assertions in KAT by

$$\{p\}\ a\ \{q\} \Leftrightarrow paq' \leq 0.$$

Using the dual of (52), we can rewrite this definition more directly as

$$\{p\}\ a\ \{q\} \Leftrightarrow p:a \leq q. \tag{81}$$

Accordingly, the inference rules of PHL can be encoded as

$$\text{Composition} \quad p\!:\!a \leq q \wedge q\!:\!b \leq r \Rightarrow p\!:\!(ab) \leq r,$$
$$\text{Conditional} \quad (pq)\!:\!a \leq r \wedge (p'q)\!:\!b \leq r \Rightarrow q\!:\!(pa + p'b) \leq r,$$
$$\text{While} \quad (pq)\!:\!a \leq q \Rightarrow q\!:\!((pa)^*p') \leq p'q,$$
$$\text{Weakening} \quad p_1 \leq p \wedge p\!:\!a \leq q \wedge q \leq q_1 \Rightarrow p_1\!:\!a \leq q_1.$$

THEOREM 10.1. *The encoded rules of* PHL *are derivable in* KAP. *Therefore* PHL *is sound with respect to this algebraic semantics.*

PROOF. (i)  (Composition)

$$p\!:\!(ab) \leq (p\!:\!a)\!:\!b \leq q\!:\!b \leq r.$$

The first step uses (33), the second one the assumption and monotonicity.

(ii)  (Conditional)

$$q\!:\!(pa + p'b) = (pq)\!:\!a + (p'q)\!:\!b \leq r + r = r.$$

(iii)  (While)

$$(pq)\!:\!a \leq q \Rightarrow q\!:\!(pa)^* \leq q \Rightarrow (q\!:\!(pa)^*)p' \leq qp' \Rightarrow q\!:\!((pa)^*p') \leq p'q.$$

The first step uses commutativity of tests and (61). The third step uses again import/export.

(iv)  (Weakening)

$$p_1\!:\!a \leq p\!:\!a \leq q \leq q_1.$$

Soundness of PHL means in our context that for every partial correctness assertion that can be proved in this calculus there is a calculation in KAD using translated statements. This follows by induction on the structure of proofs in PHL and our previous considerations.  □

□

Thus, given our domain calculus from the previous sections, soundness of PHL can be proved literally in four lines. Compared to the KAT-based approach in [Kozen 2001], we believe that our encodings and proofs in KAD are more concise and intuitive. Compared to standard set-theoretic proofs in textbooks (c.f [Apt and Olderog 1997; Loeckx and Sieber 1987]), our proof is about ten times shorter, without taking into account the fact that many logical and set-theoretic assumptions are left implicit in the textbook proofs and the proofs there are only semi-formal.

Moreover, it has already been observed in [Kozen 2001] that all Horn clauses built from partial correctness assertions in Hoare logic that are valid with respect to the standard semantics are derivable in KAT. This result holds a fortiori for KAD. PHL is too weak to derive all such formulas [Kozen 2001].

It should be noted that Hoare logic is an example where the domain operator can be completely eliminated from all expressions using (gla). Even more, all inference rules of Hoare logic can be translated into Horn clauses in KAT, where all antecedents are of the form $p = 0$. A technique for hypothesis elimination [Cohen 1994; Kozen 2001; Kozen and Smith 1996] yields decidability of this fragment.

A further investigation of PHL in KAP, notably a fully algebraic completeness proof, can be found in [Möller and Struth 2003b]. As a conclusion, we can only support [Kozen 2001] that given Kleene algebra, *the specialized syntax and deductive apparatus of Hoare logic are inessential and can be replaced by simple equational reasoning.* We also believe that KAD offers even further advantages. It allows us to combine the intuitiveness and readability of specifications in Hoare logic and imperative program semantics with the computational power of KAT. And finally, Kleene algebra offers an elegant formal calculus and a simple algebraic semantics for reasoning in and about Hoare logic.

## 11. CONCLUSION AND FURTHER WORK

We have presented equational axioms for domain and codomain for certain idempotent semirings and extended these notions to KAD. This algebraic abstraction is intended as a unified view on approaches to program analysis and development as different as PDL, KAT, B and Z. We have outlined a calculus for KAD, defined preimage and image operators and presented two applications of KAD: an algebraic reconstruction of the notions of Noethericity and the subsumption of propositional Hoare logic. These and most of the other results in this text provide the foundations and introduce the basic calculus of KAD. They are the basis for further interesting work.

On the theoretical side, expressiveness, complexity, completeness or representability of KAD have not been investigated in this text. The same holds for the apparent relation to modal algebras and in particular algebraic variants of PDL (cf. e.g. [Ehm et al. 2003]).

On the practical side, it might be interesting to continue our investigations of termination analysis and greedy algorithms [Möller and Struth 2003a]. Moreover, a combination of the two methods for total correctness reasoning seems promising. First steps in this direction with a related Kleene algebra have already been taken in [von Wright 2002]. In general, the flexibility and naturalness of KAD seems very promising for the specification and analysis of state transition systems. As often with Kleene algebra, KAD might offer an abstract, simple, elegant, uniform calculus where different specialized formalisms and complicated reasoning had to be used before.

### ACKNOWLEDGMENTS

We would like to thank Thorsten Ehm, Marcelo Frías, Hitoshi Furusawa and Dexter Kozen for discussions and helpful comments.